\documentclass[twocolumn]{aastex631}
\usepackage{mathrsfs}
\usepackage{mathcomp}
\usepackage{amssymb}
\usepackage{color}
\usepackage{soul}

\received{}
\revised{}
\accepted{}
\submitjournal{ApJ}
\shorttitle{Characterization of mm-line-emitting galaxies}
\shortauthors{Mizukoshi et al.}
\graphicspath{{./}{figures/}}

\begin{document}

\title{Physical Characterization of Serendipitously Uncovered Millimeter-wave Line-emitting Galaxies at $z\sim2.5$ behind the Local Luminous Infrared Galaxy VV~114}


\author{Shoichiro Mizukoshi}
\affiliation{Institute of Astronomy, Graduate School of Science, The University of Tokyo, 2-21-1 Osawa, Mitaka, Tokyo 181-0015, Japan}


\author[0000-0002-4052-2394]{Kotaro Kohno}
\affiliation{Institute of Astronomy, Graduate School of Science, The University of Tokyo, 2-21-1 Osawa, Mitaka, Tokyo 181-0015, Japan}
\affiliation{Research Center for the Early Universe, Graduate School of Science, The University of Tokyo, 7-3-1 Hongo, Bunkyo-ku, Tokyo 113-0033, Japan}

\author[0000-0002-1639-1515]{Fumi Egusa}
\affiliation{Institute of Astronomy, Graduate School of Science, The University of Tokyo, 2-21-1 Osawa, Mitaka, Tokyo 181-0015, Japan}

\author[0000-0001-6469-8725]{Bunyo Hatsukade}
\affiliation{Institute of Astronomy, Graduate School of Science, The University of Tokyo, 2-21-1 Osawa, Mitaka, Tokyo 181-0015, Japan}


\author[0000-0002-2933-048X]{Takeo Minezaki}
\affiliation{Institute of Astronomy, Graduate School of Science, The University of Tokyo, 2-21-1 Osawa, Mitaka, Tokyo 181-0015, Japan}

\author[0000-0002-2501-9328]{Toshiki Saito}
\affiliation{Max Planck Institute for Astronomy, K$\ddot{o}$nigstuhl 17, 69117 Heidelberg, Germany}

\author[0000-0003-4807-8117]{Yoichi Tamura}
\affiliation{Division of Particle and Astrophysical Science, Graduate School of Science, Nagoya University, Nagoya 464-8602, Japan.}

\author[0000-0002-2364-0823]{Daisuke Iono}
\affiliation{National Astronomical Observatory of Japan, 2-21-1 Osawa, Mitaka, Tokyo 181-8588, Japan}
\affiliation{The Graduate University for Advanced Studies (SOKENDAI), 2-21-1 Osawa, Mitaka, Tokyo 181-0015, Japan}

\author[0000-0003-3652-495X]{Junko Ueda}
\affiliation{National Astronomical Observatory of Japan, 2-21-1 Osawa, Mitaka, Tokyo 181-8588, Japan}

\author[0000-0003-1747-2891]{Yuichi Matsuda}
\affiliation{National Astronomical Observatory of Japan, 2-21-1 Osawa, Mitaka, Tokyo 181-8588, Japan}
\affiliation{The Graduate University for Advanced Studies (SOKENDAI), 2-21-1 Osawa, Mitaka, Tokyo 181-0015, Japan}

\author[0000-0002-8049-7525]{Ryohei Kawabe}
\affiliation{National Astronomical Observatory of Japan, 2-21-1 Osawa, Mitaka, Tokyo 181-8588, Japan}
\affiliation{The Graduate University for Advanced Studies (SOKENDAI), 2-21-1 Osawa, Mitaka, Tokyo 181-0015, Japan}

\author[0000-0002-2419-3068]{Minju M. Lee}
\affiliation{Max-Planck-Institut f$\ddot{u}$r Extraterrestrische Physik (MPE), Giessenbachstr., D-85748 Garching, Germany}

\author[0000-0001-7095-7543]{Min S. Yun}
\affiliation{Department of Astronomy, University of Massachusetts, Amherst, MA 01003, USA}

\author[0000-0002-8726-7685]{Daniel Espada}
\affiliation{Departamento de F\'{i}sica Te\'{o}rica y del Cosmos, Campus de Fuentenueva, Universidad de Granada, E18071–Granada, Spain}


\begin{abstract}

We present a detailed investigation of millimeter-wave line emitters ALMA J010748.3-173028 (ALMA-J0107a) and ALMA J010747.0-173010 (ALMA-J0107b), which were serendipitously uncovered in the background of the nearby galaxy VV~114 with spectral scan observations at $\lambda$ = 2 -- 3 mm.
Via Atacama Large Millimeter/submillimeter Array (ALMA) detection of CO(4--3), CO(3--2), and [C\,{\sc i}](1--0) lines for both sources, their spectroscopic redshifts are unambiguously determined to be 
$z= 2.4666\pm0.0002$ and $z=2.3100\pm0.0002$, respectively. 
We obtain the apparent molecular gas masses $M_{\rm gas}$ of 
these two line emitters {from [C\,{\sc i}] line fluxes} as $(11.2 \pm 3.1) \times 10^{10} M_\odot$ and $(4.2 \pm 1.2) \times 10^{10} M_\odot$, respectively.
The observed CO(4--3) velocity field of ALMA-J0107a exhibits a clear velocity gradient across the CO disk, and we find that ALMA-J0107a is characterized by an inclined
rotating disk with a significant turbulence, that is, a deprojected maximum rotation velocity to velocity dispersion ratio $v_{\rm max}/\sigma_{v}$ of $1.3 \pm 0.3$.
We find that the dynamical mass of ALMA-J0107a within the CO-emitting disk computed from the derived kinetic parameters, $(1.1 \pm 0.2) \times 10^{10}\ M_\odot$, is an order of magnitude smaller than the { molecular gas mass derived from dust continuum emission}, $(3.2\pm1.6)\times10^{11}\ M_{\odot}$.
We suggest {this  source is magnified by} a gravitational lens with a magnification of $\mu \ga10$, 
which is consistent with the measured offset
from the empirical correlation between CO-line luminosity and width.

\end{abstract}

\section{Introduction} \label{sec:intro}

In our universe, galaxies form stars most actively at $z=1-3$ \citep{Hopkins06, Madau14}, and their molecular gas content is a key parameter because stars are formed in molecular gas. Therefore, extensive observations of rotational CO lines, which have been established as a useful measure of cold molecular gas mass $M_{\rm gas}$ \citep[e.g., ][]{Bolatto13}, have been made for various samples of galaxies that are pre-selected based on their physical properties, such as stellar mass ($M_\star$) and star formation rate (SFR), \citep[e.g.,][]{Tacconi20}.
This approach has successfully revealed the evolution of molecular gas components by measuring the molecular gas fraction $f_{\rm gas} \equiv M_{\rm gas}/(M_{\rm gas}+M_\star)$ in galaxies across cosmic time \citep[e.g.,][and references therein]{Tacconi18}. 
Despite its success, it is also necessary to conduct a {\it blind} search of CO-line-emitting galaxies without any priors. This can be accomplished by unbiased spectral scan observations of a region of the sky, 
{which often target
known deep fields}
such as the Hubble Ultra Deep Field (HUDF), where rich multi-wavelength datasets are available.
This ``deep-field scanning approach'' is capable of uncovering galaxies that were not present in standard optical/near-infrared deep surveys, and thus is considered less biased than the pointed approach { \citep{Carilli13,Tacconi20} }. Following the pioneering spectral scan observations of the Hubble Deep Field North (HDF-N) using the IRAM Plateau de Bure Interferometer \citep{Decarli14}, the Atacama Large Millimeter/submillimeter Array (ALMA) has been exploited to conduct spectral scan observations of the HUDF \citep[e.g., ][]{Walter16,Aravena19}, SSA22 \citep{Hayatsu17,Hayatsu19}, and lensing clusters \citep[e.g., ][]{Yamaguchi17,Gonzalez17} to uncover millimeter-wave line-emitting galaxies and constrain CO-line luminosity functions as a function of redshift and, therefore, the cosmic molecular gas mass density evolution \citep[e.g.,][]{Decarli20}.

In addition to these dedicated spectral scan observations of deep fields, there are mounting examples of serendipitous detection of millimeter-wave line emitters.
They were detected using ALMA and the NOrthern Extended Millimeter Array (NOEMA), {and in the data from the second Plateau de Bure High-z Blue- Sequence Survey (PHIBSS2)}, within a field of view (FoV) of, for example, a nearby galaxy \citep[e.g., ][]{Tamura14} and high-$z$ source { \citep[e.g., ][]{Swinbank12,Gowardhan17,Wardlow18,Lenkic20}. }
Currently, the nature of such serendipitously uncovered millimeter-wave line emitters remains unexplored given the very limited number of such sources, but it is important to characterize the known line emitters because we can learn the types of galaxies that can be selected from monotonically increasing numbers of spectral cubes in the ALMA science archive over time. 

Here, we report a detailed investigation of millimeter-wave line emitters ALMA J010748.3-173028 (ALMA-J0107a) and ALMA J010747.0-173010 (ALMA-J0107b), which were serendipitously uncovered around the nearby galaxy VV~114 with spectral scan observations at $\lambda$ = 2 -- 3 mm. 
We display the positions of ALMA-J0107a and ALMA-J0107b in Figure \ref{fig:HST} (top), in which an HST $I$-band image of VV~114 is shown. 
As VV~114 is one of the best-studied archetypical luminous infrared galaxies (LIRGs) in the local region \citep[e.g.,][]{Iono13,Saito15,Saito17}, a number of spectral scan observations have been conducted using ALMA. 
The discovery of ALMA-J0107a was first reported by \cite{Tamura14}, based on a single line detection in ALMA band 3. 
Although the multi-wavelength counterpart identification at the position of ALMA-J0107a favors a CO(3--2) line at $z=2.467$, the proximity of ALMA-J0107a to the local LIRG VV~114 ($\sim10\arcsec$ from the eastern nucleus of VV~114) hampers reliable photometric constraints at near-to-far-infrared bands and therefore requires other  transitions of CO to obtain an unambiguous spectroscopic redshift.
{\cite{Tamura14} also found the hard X-ray source at the position of ALMA-J0107a in} \it{}Chandra\rm{}{/ACIS-I data (see Figure \ref{fig:HST} (a)), which suggested the presence of an active galactic nucleus (AGN).}
{ALMA-J0107b is also serendipitously detected in the line scan of the VV~114 field at $(\alpha, \delta)_{\rm{J2000}}=(01^{\rm{h}}07^{\rm{m}}46^{\rm{s}}.99, -17\tcdegree30'10\farcs09)$, and we report it in this paper.} 
{Figure \ref{fig:HST}(a and b) shows multiwavelength images of both ALMA-J0107a and ALMA-J0107b.
It contains three} \it{}Spitzer\rm{}{/IRAC band images and the} \it{}Chandra\rm{}{/ACIS-I image.}

The remainder of this paper is organized as follows: The reduced ALMA data and reduction procedures are described in Section \ref{sec:data analysis} with the derived line spectra. The physical quantities derived from the observed lines and continuum emissions are summarized in Section \ref{sec:Results}. Section \ref{sec:kinematic modeling} is devoted to kinematic modeling of the observed CO velocity field of ALMA-J0107a. After discussing the nature of the millimeter-wave line emitters in Section \ref{sec:Discussion}, we summarize our findings in Section \ref{sec:conclusion}.
We assume a $\Lambda$CDM cosmology 
with $\Omega_{\mathrm{m}}=0.3, \Omega_{\mathrm{\Lambda}}=0.7$, and $H_0=68$ km s$^{-1}$ Mpc$^{-1}$.

\section{Data analysis} \label{sec:data analysis}

\subsection{Selection, reduction, and imaging}
\label{subsec:2.1}

We analyzed the ALMA data listed in Table \ref{tab:ALMAdata}, which targeted VV~114 and include J0107a and J0107b within the FoV. 
{The data sets of bands 3 and 4 were selected based on the frequency range, in which some CO and [C\,{\sc i}] emission lines should be included if the redshift estimate of $z=2.467$ \citep{Tamura14} was correct.
The data sets of bands 6 and 7 were used to measure the dust continuum emissions. We selected these data sets based on the relative position of VV114 in the FoV in order to detect our targets, which is often in the edge of the FoV.  
We also selected the data sets with relatively long integration time, more than 1000 s, to achieve high signal-to-noise ratio (S/N). }
We used channels without emission lines to analyze continuum emission.
We conducted standard calibration and imaging using the Common Astronomy Software Applications package \citep[CASA versions 5.1.0 and 5.4.0;][]{McMullin07}. 

For imaging, we used the CASA task {\sf tclean} with a parameter {\sf threshold} of 1--$3\ \sigma$. 
Briggs weighting with a robust parameter of {\sf robust = 0.5} was adopted for the band 3 and 4 data, while {\sf robust = 2.0} was adopted for band 6 and 7 data, for which the beam size was significantly smaller than for bands 3 and 4.
{The data of the [C\,{\sc i}] line (see Section \ref{subsec:line_identification}) and dust continuum (see Section \ref{subsec:continuum_data}) are strongly affected by the emission from VV114. 
Therefore, we set a mask as a box around VV~114 for these data to effectively remove the side lobes.}

\tabletypesize{\scriptsize}
\begin{deluxetable*}{ccccccc}
\tablecaption{The details of ALMA archive data we used in this study.}
\label{tab:ALMAdata}
\tablewidth{0pt}
\tablehead{
\colhead{Project ID} & \colhead{Band} &Max baseline length& \colhead{Frequency (GHz)}& \colhead{Integration time (s)} & \colhead{Target $^\mathrm{a}$}& \colhead{Main use $^\mathrm{b}$}
}
\startdata
2013.1.01057.S&3&$650.3\,$m&84.08-87.79 / 97.91-99.79&2268.0&a&CO(3--2), continuum  \\
2013.1.01057.S&3&$650.3\,$m&87.81-91.56 / 99.81-103.55&1360.8&a&CO(3--2), continuum  \\
2013.1.01057.S&3&$783.5\,$m&91.56-95.31 / 103.56-107.31 &2721.600&b&CO(3--2), continuum  \\
2013.1.01057.S&4&$538.9\,$m&130.74-134.49 / 142.74-146.49&1149.120&a&CO(4--3), continuum  \\
2013.1.01057.S&4&$538.9\,$m&138.24-141.99 / 150.24-153.99&1149.120&b&CO(4--3), continuum  \\
2013.1.01057.S&4&$538.9\,$m&126.99-130.74 / 138.99-142.74&1149.120&a&[C\,{\sc i}](1--0), continuum  \\
2013.1.01057.S&4&$538.9\,$m&134.49-138.24 / 146.49-150.24&1149.120&b&[C\,{\sc i}](1--0), continuum  \\
2015.1.00973.S&6&$641.5\,$m&245.09-248.93 / 259.44-263.14&1814.400&a / b&continuum  \\
2015.1.00902.S&6&$867.2\,$m&211.87-215.07 / 226.05-228.23&2721.600&a / b&continuum  \\
2013.1.00740.S&7&$1.6\,$km&325.67-329.50 / 337.67-341.49&1332.197&a / b&continuum  \\
2013.1.00740.S&7&$1.6\,$km&334.89-338.70 / 346.87-350.49&2124.645&a / b&continuum  \\
\enddata
\tablecomments{\\$^\mathrm{a}$ target a represents J0107a and target b represents J0107b. \\
$^\mathrm{b}$ Suggested emission lines are detected only with the target suggested in this table. 
}
\end{deluxetable*}

\tabletypesize{\scriptsize}
\begin{deluxetable*}{ccccccc}
\tablecaption{Summary of the analysis of detected emission lines in this study.}
\label{tab:lineresults}
\tablewidth{0pt}
\tablehead{
\colhead{}&\colhead{J0107a}&\colhead{}&\colhead{}&\colhead{J0107b}&\colhead{}&\colhead{}\\
\colhead{} & \colhead{CO(4--3)} & \colhead{CO(3--2)} & \colhead{[C\,{\sc i}](1--0)}& \colhead{CO(4--3)} & \colhead{CO(3-2)}&\colhead{[C\,{\sc i}](1--0)}
}
\startdata
$\nu_{\rm{obs}}$ (GHz)&132.99&99.749&141.97 &139.29&104.47&148.69 \\ 
{Line peak (mJy beam$^{-1}$)}  &$2.29\pm0.11$&$1.88\pm12$&$0.93\pm0.13$ &$1.77\pm0.10$&$1.06\pm0.12$&$0.38\pm0.05$ \\  
rms (mJy beam$^{-1}$)&$0.54$&$0.42$&$0.46$&$0.33$&$0.37$&$0.32$ \\beam size &$1\farcs23\times1\farcs02$&$1\farcs24\times1\farcs03$&$1\farcs37\times1\farcs00$&$1\farcs22\times0\farcs955$&$1\farcs16\times0\farcs916$&$1\farcs02\times0\farcs870$\\
deconvolved source size &$1\farcs3\times0\farcs7$&$1\farcs8\times1\farcs0$&$1\farcs1\times0\farcs3$&$0\farcs7\times0\farcs5$&$0\farcs7\times0\farcs4$&$1\farcs0\times0\farcs3$ \\
$S_{\rm{obs}}$ (Jy\ km s$^{-1}$)  &$3.0\pm0.2$&$2.2\pm0.2$&$1.2\pm0.2$&$1.3\pm0.1$&$0.95\pm0.1$&$0.51\pm0.1$  \\ 
Luminosity ($10^{10}$ K km s$^{-1}$ pc$^2$)
&$5.7\pm0.4$&$7.2\pm0.5$&$2.0\pm0.3$&$2.1\pm0.2$&$2.8\pm0.3$&$0.75\pm0.1$\\
Line FWHM (km s$^{-1}$)  &$193\pm11$&$165\pm13$&$217\pm34$ &$190\pm13$&$164\pm21$&$176\pm27$ \\  
Peak S/N in channel maps &$\sim\,20$&$\sim\,15$&$\sim\,10$ &$\sim11$&$\sim9$ &$\sim3$ \\ 
\enddata
\tablecomments{ 
 {The line peaks are those of line spectra in Figure \ref{fig:label_2}.} The source size and velocity-integrated flux density ($S_{\rm{obs}}$) are measured with the CASA task \sf{}imfit\rm{} and we applied primary beam correction for $S_{\rm{obs}}$. 
The velocity resolution is 20 km s$^{-1}$ for the data of CO(4--3) of J0107a and 50 km s$^{-1}$ for other data. Source sizes have 5--10 \% errors for each axis. {The peak S/N is the approximate ratio of line peak in Figure \ref{fig:all_spectrum} and rms noise for each line.}}
\end{deluxetable*}

\subsection{Line identification}
\label{subsec:line_identification}

\begin{figure*}
\gridline{\fig{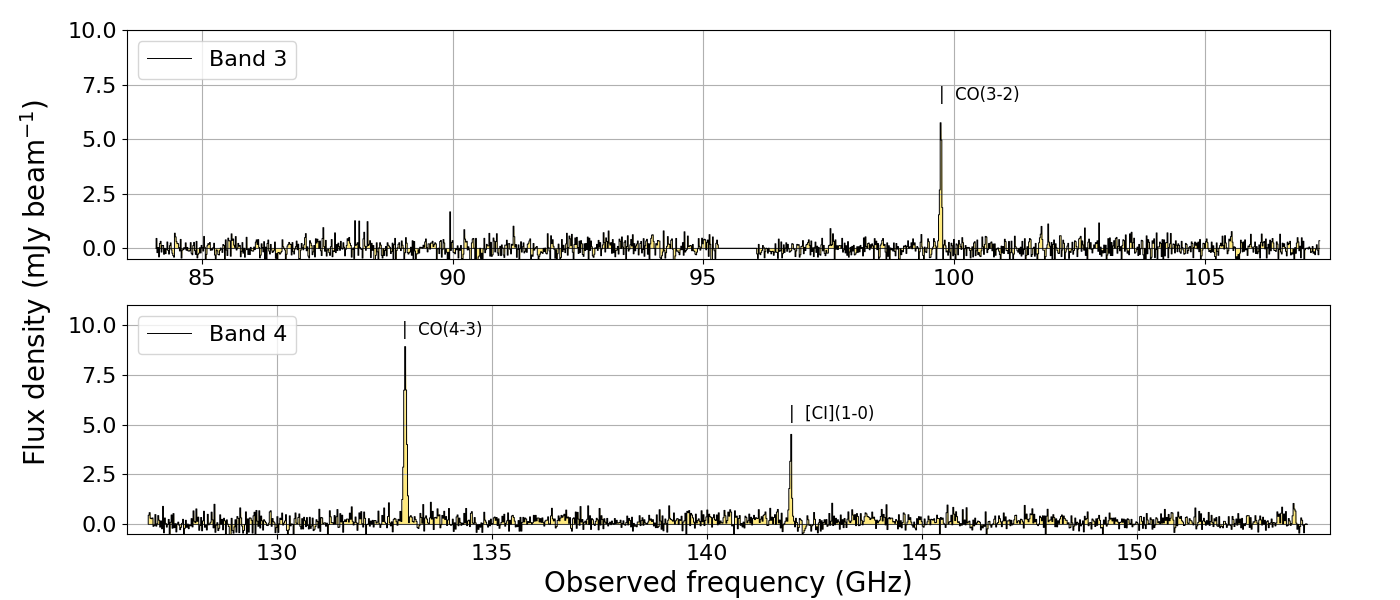}{0.9\textwidth}{(a)}
		  }
\gridline{\fig{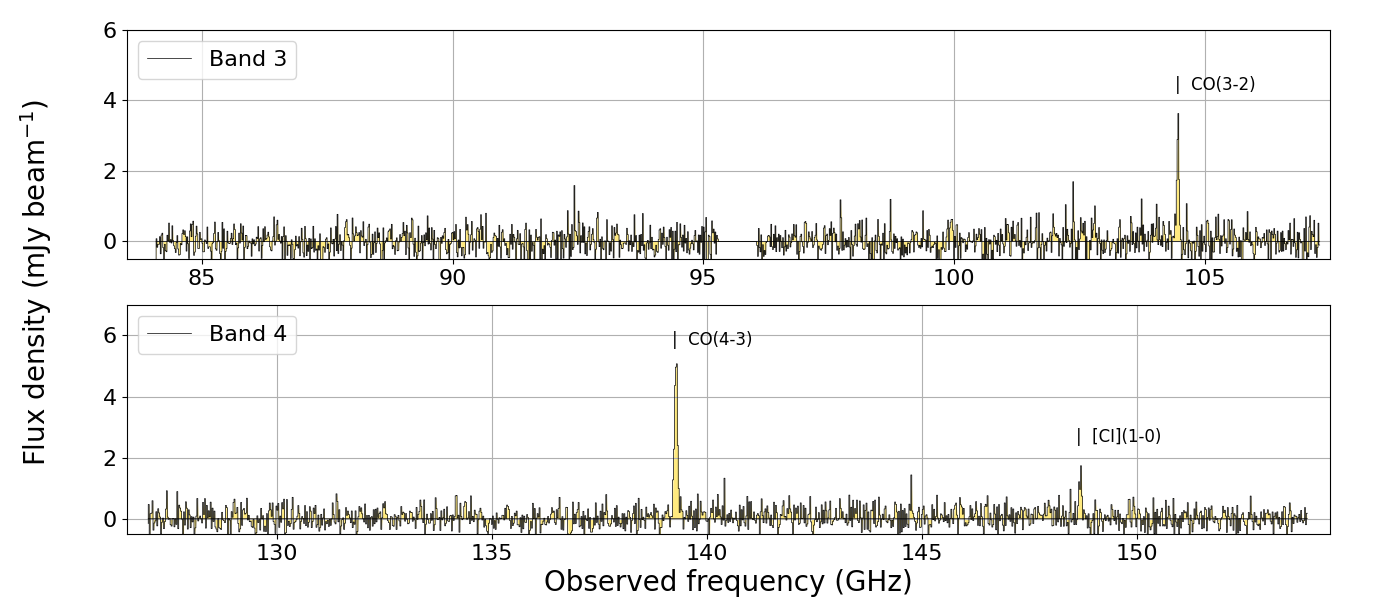}{0.9\textwidth}{(b)}
          }
\caption{
The band 3 and 4 spectra for J0107a (a) and J0107b (b). The gaps seen in the band 3 spectra are due to ones between two spectral windows. 
}
\label{fig:all_spectrum}
\end{figure*}

Figure \ref{fig:all_spectrum} shows the primary-beam-corrected spectra of J0107a and J0107b over the entire range of band 3 and 4 data listed in Table \ref{tab:ALMAdata}. 
These spectra were obtained from a single pixel ($0\farcs35\times0\farcs35$) at the line peak (the same position for all lines within uncertainties) to make these line peaks clear and easy to identify. 
{Note that the peak spectra were used only for line identification and redshift determination, but not for measuring line width and flux.}

{With the use of Gaussian fittings for the peak line spectra, we identify three redshifted emission lines both in the spectra of J0107a and J0107b, as CO(3--2), CO(4--3), and [C\,{\sc i}]($^3$P$_1$--$^3$P$_0$) ([C\,{\sc i}](1--0) hereafter) at $z(\rm{J0107a})=2.4666\pm0.0002$ and $z(\rm{J0107b})=2.3100\pm0.0002$, respectively. }
The detection of multiple emission lines yields unambiguous redshifts of the two sources, and
this confirms the line identification of \cite{Tamura14}, which is based on a photometric redshift analysis using infrared-to-radio data. 
The angular diameters corresponding to $1''$ at these redshifts are 8.3 kpc and 8.4 kpc, respectively.

The zoom-in spectrum of each detected line with the best-fit Gaussian profiles is shown in Figure \ref{fig:label_2}.
{Each spectrum in this figure was made by taking an aperture of $\sim3''\times3''$ square centered at the peak position.}
{The line width of each line is measured as the full-width-at-half-maximum (FWHM) of these Gaussian profiles.
}

{We made channel maps for each line to measure line properties and to create moment maps.} 
We set the velocity resolution of the channel map which includes CO(4--3) of J0107a to be 20 km s$^{-1}$, and 50 km s$^{-1}$ for other maps.
{The derived physical properties, achieved peak S/N in channel maps, and typical noise level for each line are listed in Table \ref{tab:lineresults}.} 

To create all moment maps {(0th, 1st, and 2nd)}, we 
included approximately $\pm 150$ km\,s$^{-1}$ around the line center.
In addition, to create the 1st and 2nd moment maps, we set a masking threshold with a range of 2 -- 4 $\sigma$ of these line data.
We arbitrarily set these clipping thresholds for each data cube in order to make these maps clear.
{We do not adopt this clipping procedure for 0th moment maps, hence the total fluxes of all lines are correctly measured. In fact, the total fluxes derived from moment maps are well consistent with those from integration of the Gaussian fitting for line spectra (Figure \ref{fig:label_2}) for all emission lines. }
{We measure the integrated line fluxes $S_{\rm{obs}}$ and beam deconvolved source sizes of each line using the CASA task} \sf{}imfit\rm{} {for 0th moment maps with the default setting.}

Moment maps for the three emission lines are presented in Figures \ref{fig:J0107a} and \ref{fig:J0107b}.

\subsection{Continuum measurement}
\label{subsec:continuum_data}

\begin{figure*}
\gridline{\fig{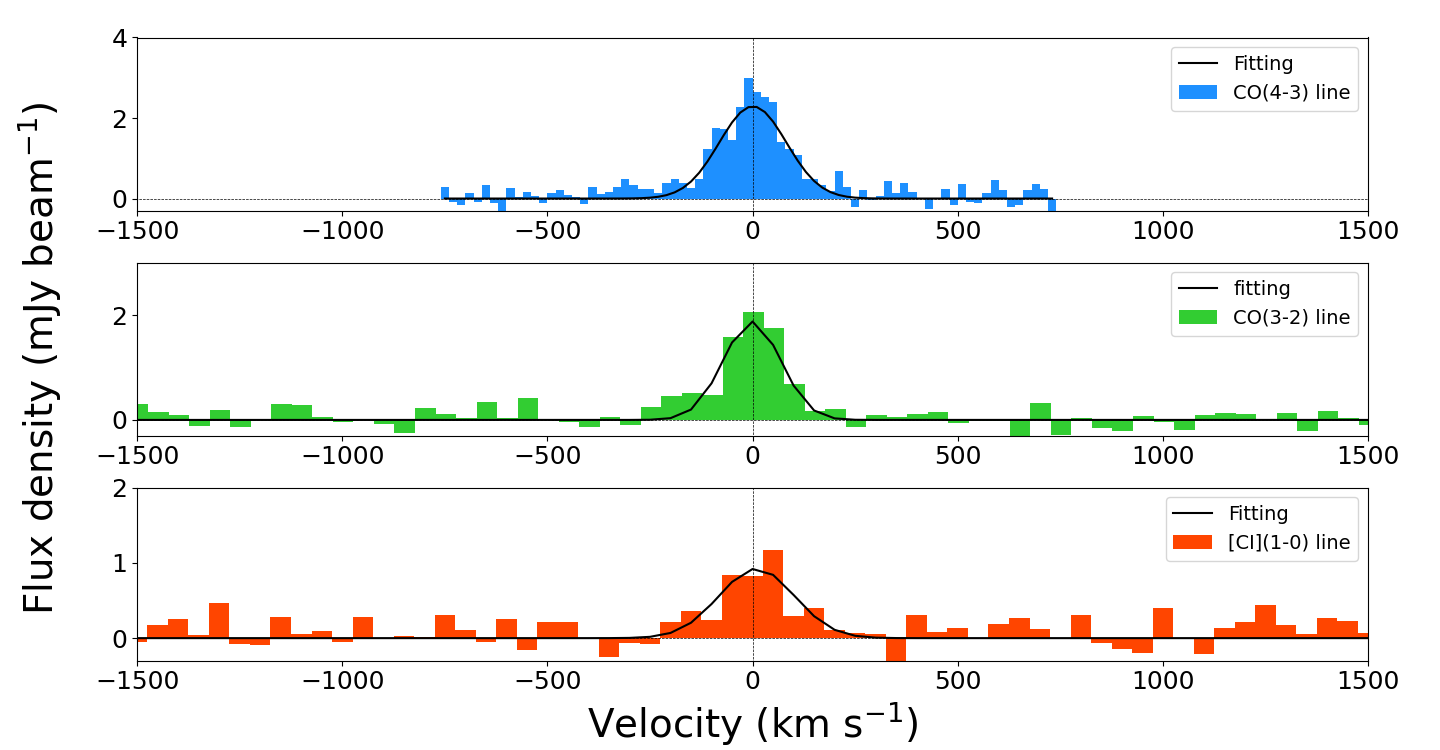}{0.45\textwidth}{(a)}
          \fig{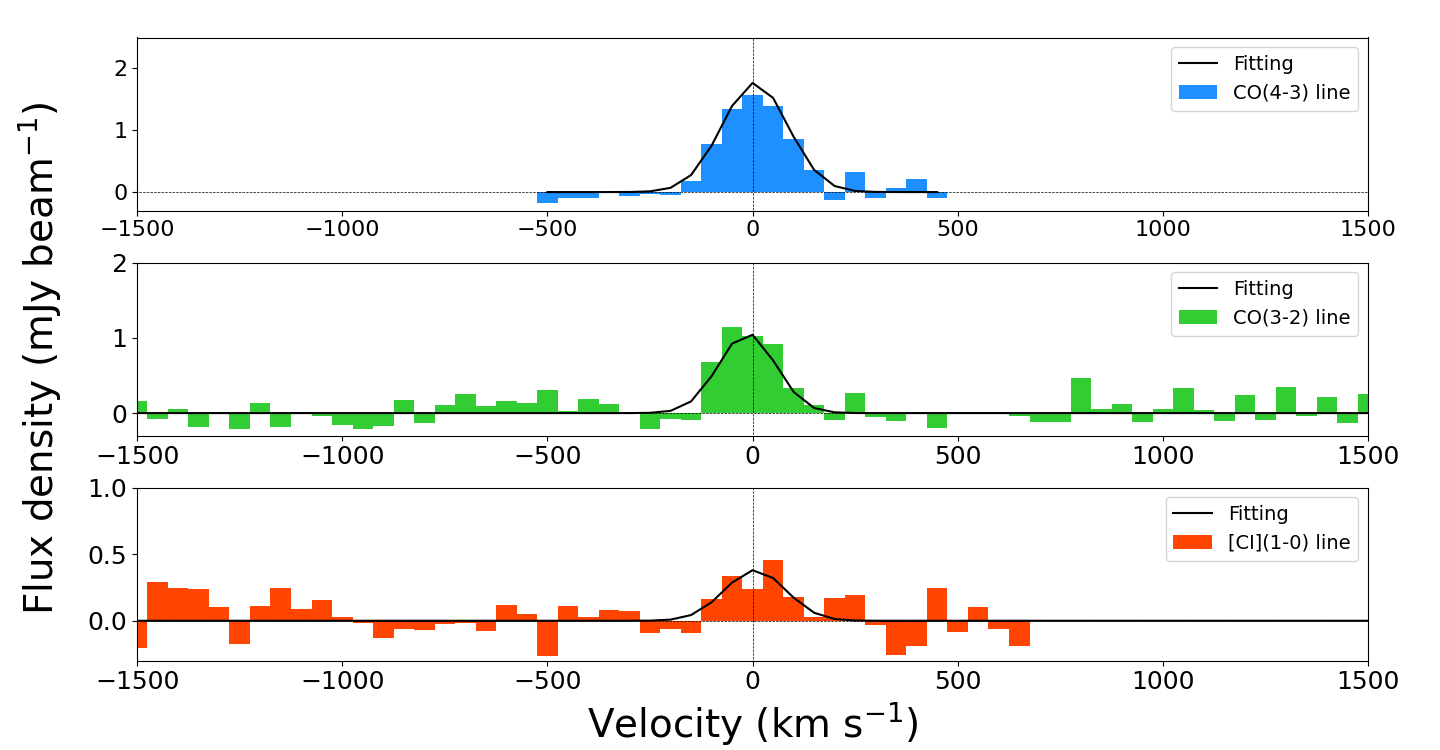}{0.45\textwidth}{(b)}
          }
\caption{The three spectral lines, CO(4--3), CO(3--2), [C\,{\sc i}](1--0), in each of J0107a (a) and J0107b (b). The black solid curve drawn in each plot is the result of Gaussian fitting to each line. The velocity resolution of data is 20 km s$^{-1}$ for CO(4--3) of J0107a and 50 km s$^{-1}$ for the others. }
\label{fig:label_2}
\end{figure*}

\begin{figure*}
\gridline{\fig{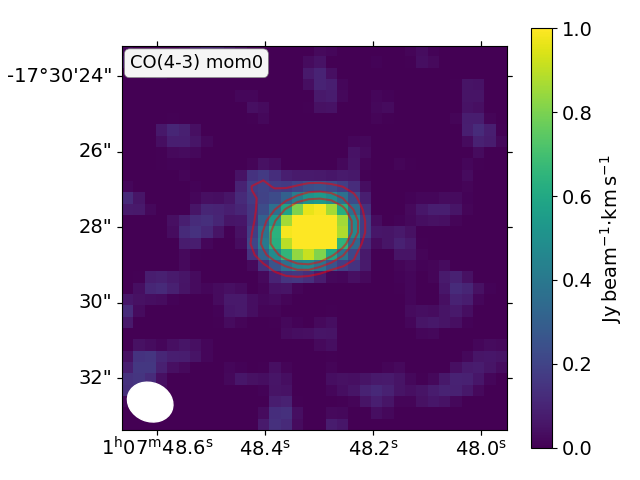}{0.25\textwidth}{(a1)}
          \fig{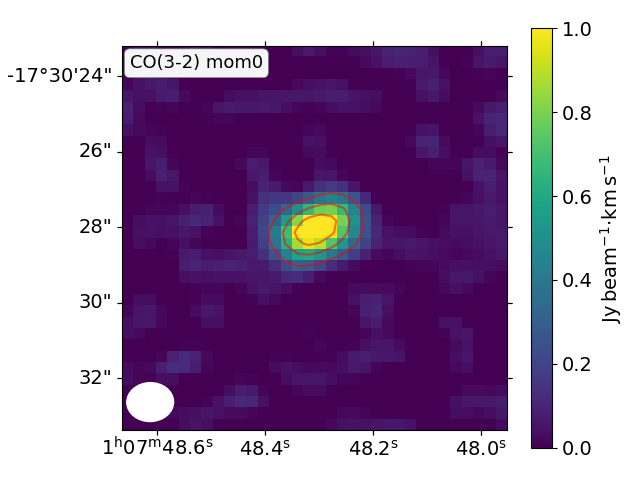}{0.25\textwidth}{(a2)}
          \fig{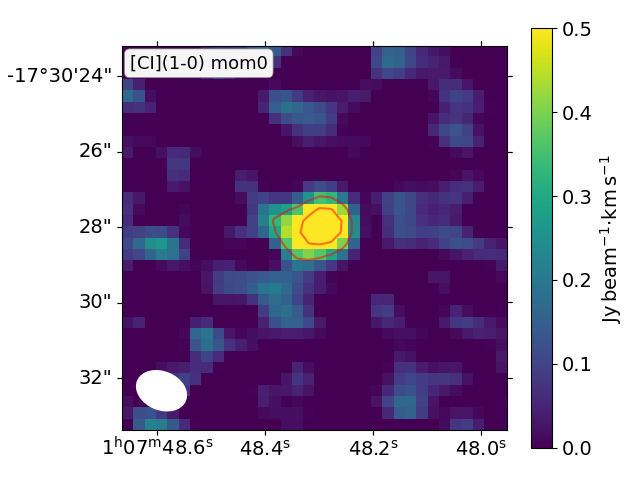}{0.25\textwidth}{(a3)}
          }
          \gridline{\fig{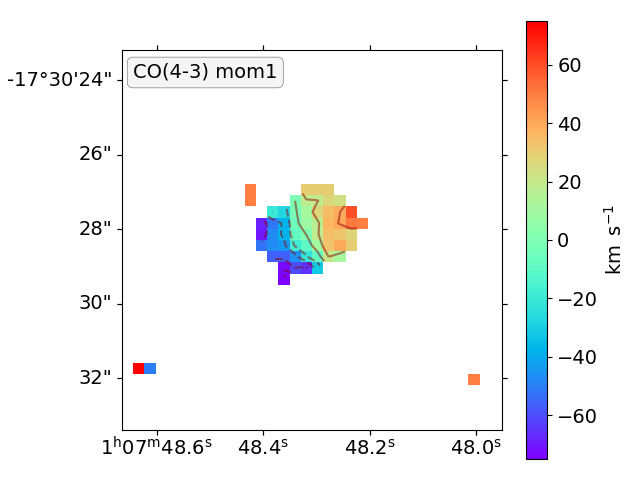}{0.25\textwidth}{(a4)}
          \fig{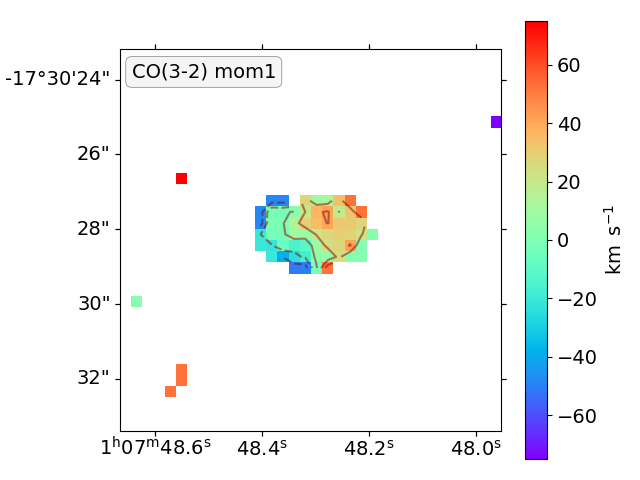}{0.25\textwidth}{(a5)}
          \fig{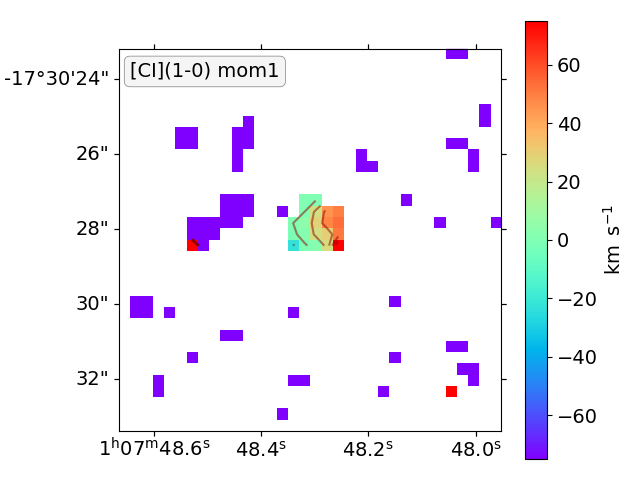}{0.25\textwidth}{(a6)}
          }
           \gridline{\fig{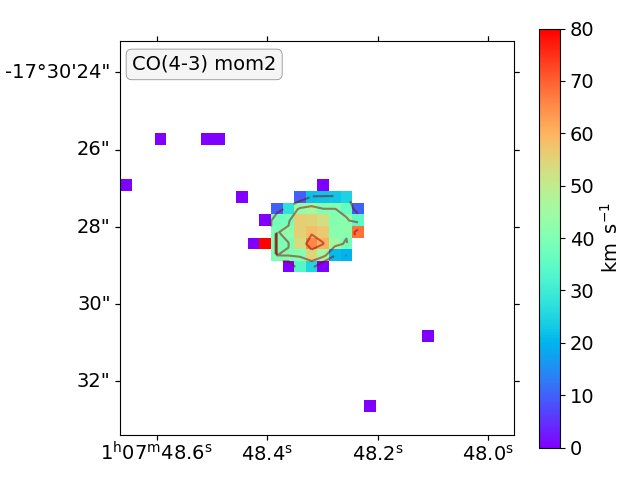}{0.25\textwidth}{(a7)}
          \fig{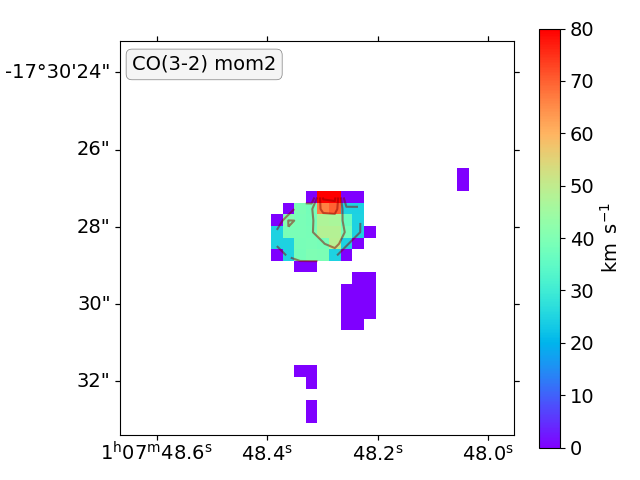}{0.25\textwidth}{(a8)}
          \fig{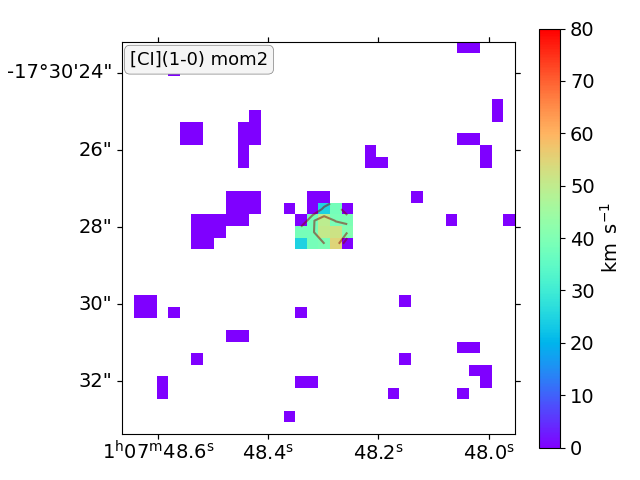}{0.25\textwidth}{(a9)}
          }
          \gridline{\fig{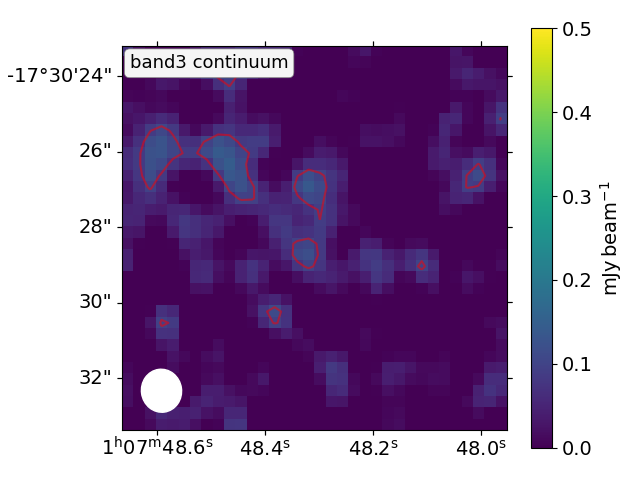}{0.25\textwidth}{(a10)}
          \fig{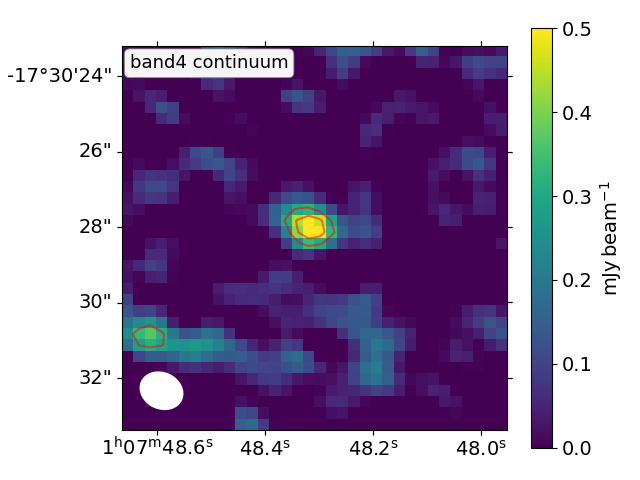}{0.25\textwidth}{(a11)}
          \fig{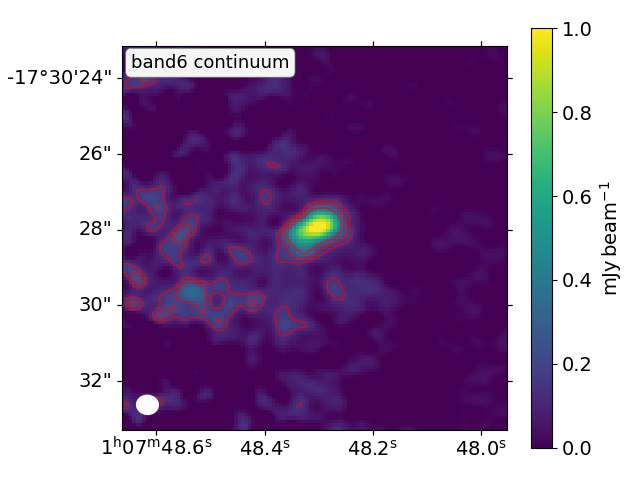}{0.25\textwidth}{(a12)}
          \fig{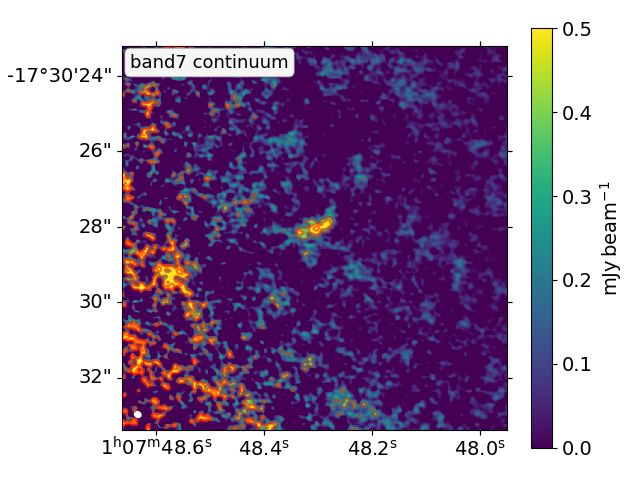}{0.25\textwidth}{(a13)}
          }
 \caption{
 CO, [C\,{\sc i}], and continuum images of J0107a. Each panel depicts a $10\arcsec \times 10\arcsec$ region centered on the CO(4--3) peak position. (a1)(a2)(a3): Velocity-integrated intensity images of CO(4--3), CO(3--2), and [C\,{\sc i}](1--0) lines. Contour levels are 10, 20, and 30 $\sigma$, where 1 $\sigma$ = 0.018, 0.031, and 0.031 Jy beam$^{-1}$ km s$^{-1}$ for CO(4--3), CO(3--2), and [C\,{\sc i}](1--0) lines, respectively. (a4)(a5)(a6): Intensity-weighted mean radial velocity images of CO(4--3), CO(3--2), and [C\,{\sc i}](1--0) lines. Contour levels are -60, -40, -20, 0, 20, and 40 km s$^{-1}$ for CO(4--3), { -40, -20, 0, 20, and 40 km s$^{-1}$ for CO(3--2), and -20, 0, 20, and 40 km s$^{-1}$ for [C\,{\sc i}](1--0)}. Negative contours are dashed. (a7)(a8)(a9): Intensity-weighted velocity dispersion images of CO(4--3), CO(3--2), and [C\,{\sc i}](1--0) lines. Contour levels are 20, 40, and 60 km s$^{-1}$ for CO(4--3) { and CO(3--2), and 20 and 40 km s$^{-1}$ for [C\,{\sc i}](1--0)}. (a10)(a11)(a12)(a13): Continuum images of 3.0 mm (band 3), 2.2 mm (band 4), 1.3 mm (band 6), and 0.89 mm (band 7). Contour levels are 1 $\sigma$ (band 3), 2 and 3 $\sigma$ (band 4), and 2, 3, and 5 $\sigma$ (bands 6 and 7). Each noise level is given in Table \ref{tab:continuum} (rms in mJy beam$^{-1}$). The band 7 continuum image (a13) has relatively large noise in the left side because J0107a is near the east edge of the field.
 }
\label{fig:J0107a}
         \end{figure*}

\begin{figure*}        
\gridline{\fig{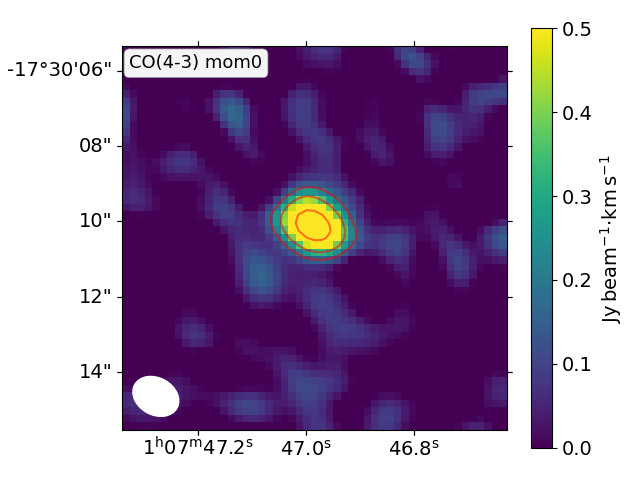}{0.25\textwidth}{(b1)}
          \fig{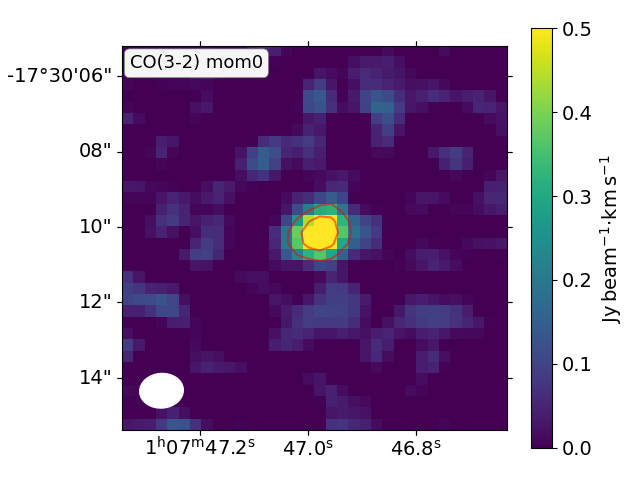}{0.25\textwidth}{(b2)}
          \fig{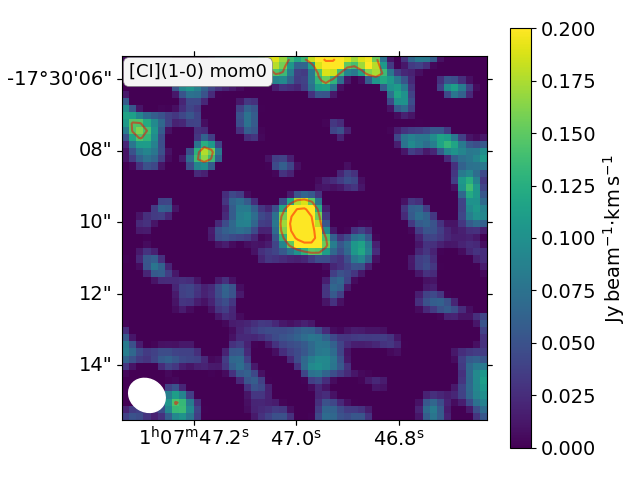}{0.25\textwidth}{(b3)}
          }
           \gridline{\fig{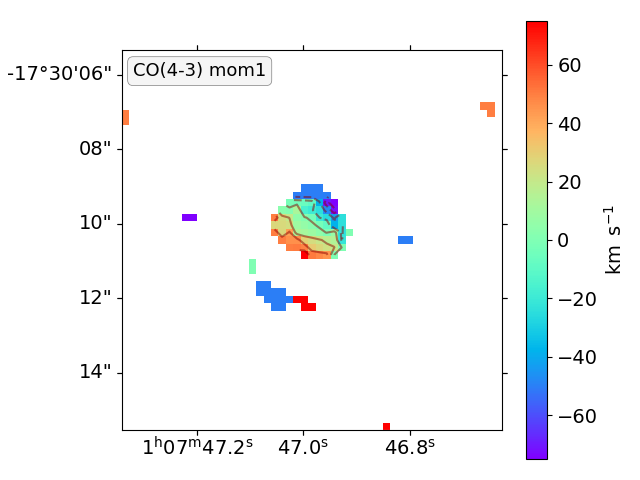}{0.25\textwidth}{(b4)}
          \fig{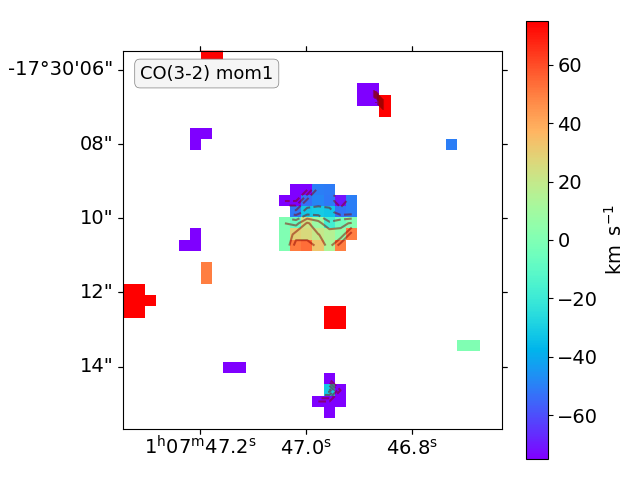}{0.25\textwidth}{(b5)}
          \fig{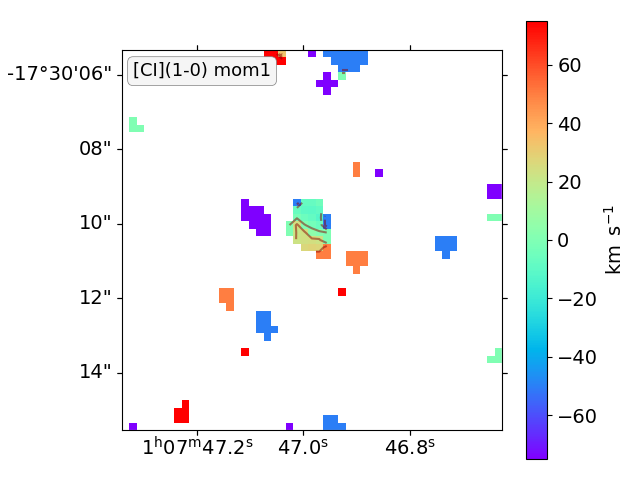}{0.25\textwidth}{(b6)}
          }
           \gridline{\fig{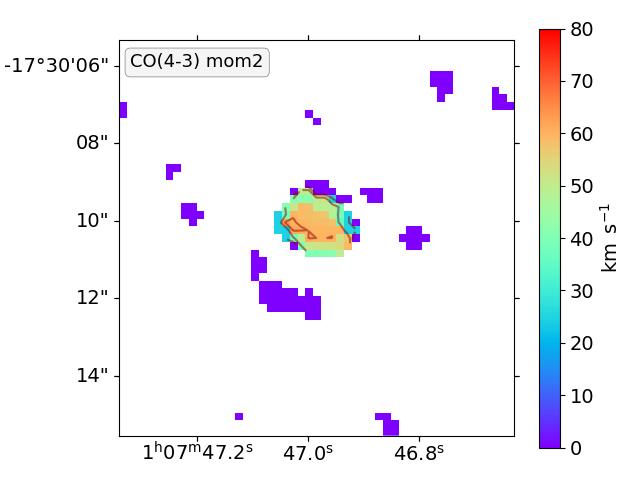}{0.25\textwidth}{(b7)}
          \fig{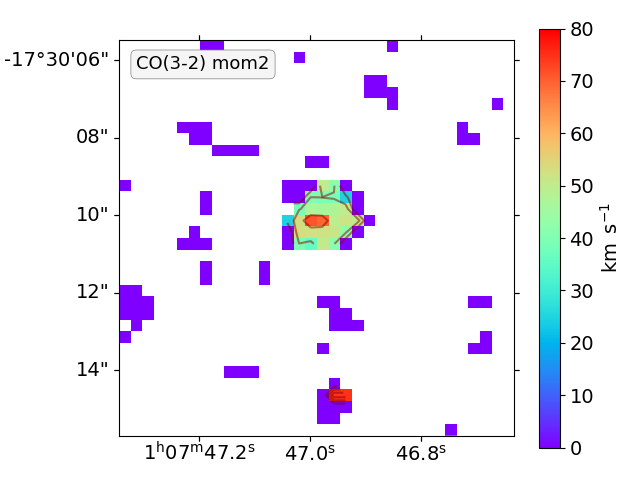}{0.25\textwidth}{(b8)}
          \fig{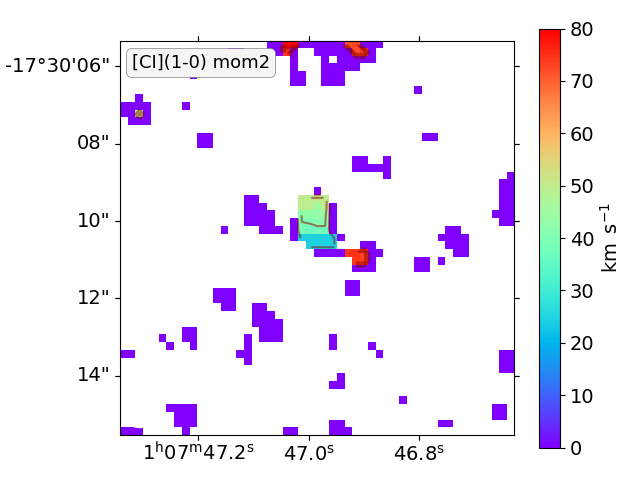}{0.25\textwidth}{(b9)}
          }
          \gridline{\fig{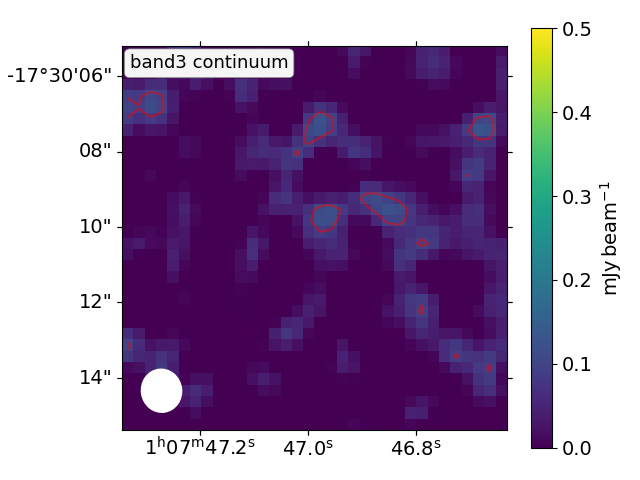}{0.25\textwidth}{(b10)}
          \fig{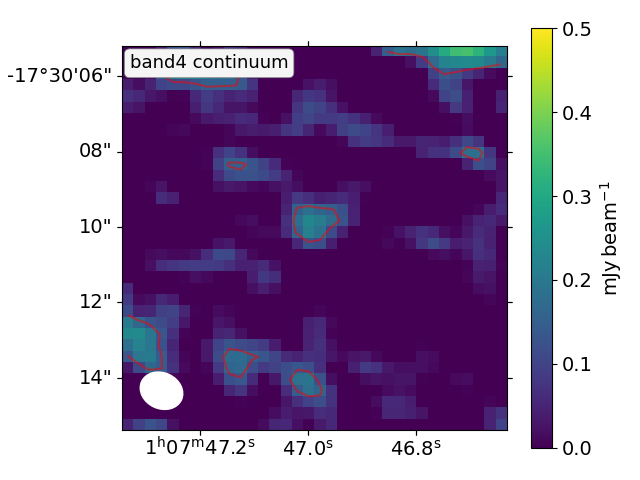}{0.25\textwidth}{(b11)}
          \fig{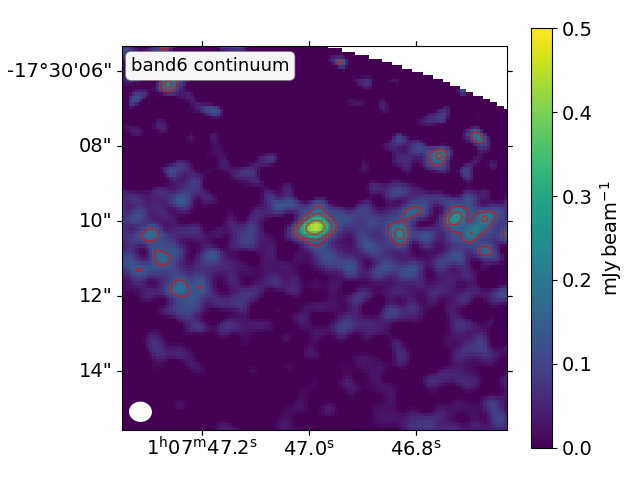}{0.25\textwidth}{(b12)}
          \fig{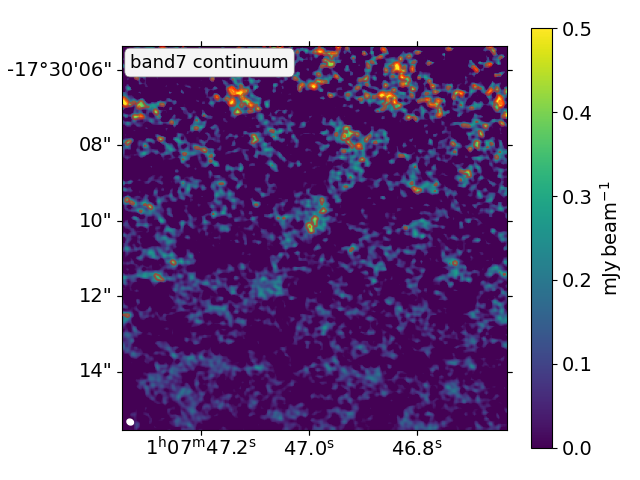}{0.25\textwidth}{(b13)}
          }
          \caption{The same as Figure \ref{fig:J0107a} but for J0107b. Contour levels in (b1), (b2), and (b3) are 5, 10, 20 $\sigma$, where 1 $\sigma$ = 0.035, 0.046, and 0.028 Jy beam$^{-1}$ km s$^{-1}$ for CO(4--3), CO(3--2), and [C\,{\sc i}](1--0) lines, respectively. Contour levels in other maps are the same as those in Figure \ref{fig:J0107a}. The band 6 (b12) and 7 (b13) continuum images are near the edges of the observed fields.
}
\label{fig:J0107b}
\end{figure*}

We analyzed the entire ALMA band 6, 7 and 3, 4 data 
using the line-free channels 
to obtain the dust continuum properties (Table \ref{tab:continuum}).
We obtain flux values using the CASA task \sf{}imfit\rm{} by setting the same area as we set to obtain line spectra in Figure \ref{fig:label_2}. The rms noise levels are measured using the CASA task \sf{}imstat\rm{} with algorithm \sf{}biweight\rm{}{ over the entire FoV}.
\rm{}Here, we did not detect the continuum of J0107a in band 3 and {of J0107b in bands 3 and 4}; hence, we present the 3--$\sigma$ upper limits for flux values in these bands. 
The intensity maps for the continuum {in these bands} are presented in Figures \ref{fig:J0107a} and \ref{fig:J0107b}. 
Because both J0107a and J0107b are located close to the edge of the band 7 FoV, these maps appear noisy. 
However, the noise levels in the vicinity of the sources are consistent with those measured over the entire field, suggesting no significant effect of the position on the map.


\begin{figure*}
\gridline{\fig{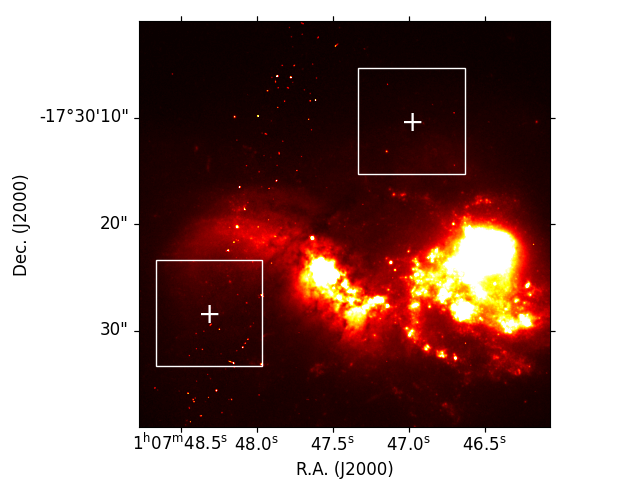}{0.6\textwidth}{}
}
\gridline{\fig{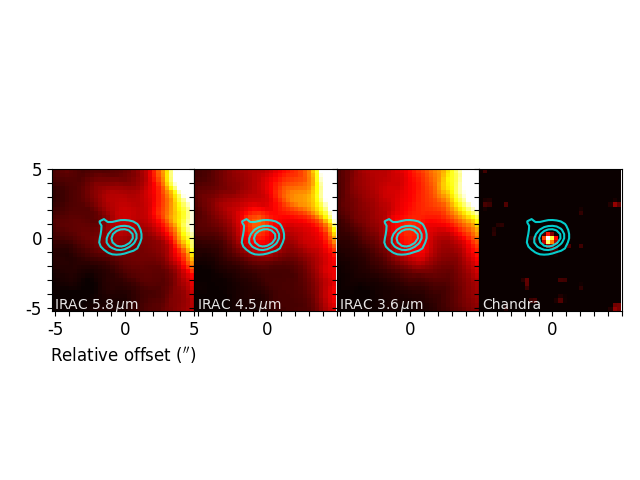}{0.5\textwidth}{(a)}
          \fig{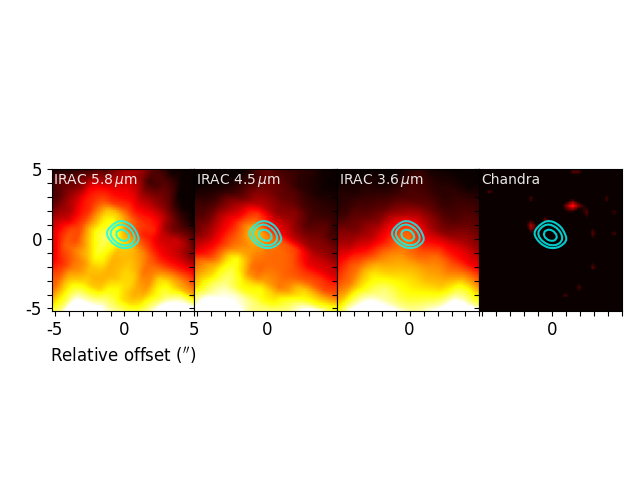}{0.5\textwidth}{(b)}
          }
\caption{\normalsize{}Top: HST image of  VV114 and the periphery, which is taken in the $I$-band (filter [F814W]). 
The two white boxes indicate the regions for which ALMA images for two sources are presented in Figures \ref{fig:J0107a} and \ref{fig:J0107b}; the east one is for J0107a, and the north one is for J0107b. The white crosses in these boxes indicate the positions of J0107a and J0107b, respectively.
{Bottom: $10''\times10''$ multiwavelength (infrared to X-ray) images of J0107a (a) and J0107b (b). 
Contours show the CO(4--3) integrated intensities and are drawn at 10, 30, and 50 $\sigma$ for J0107a, and 5, 10, and 20 $\sigma$ for J0107b. The $\sigma$ for each source are $\sigma=0.018\ \mathrm{Jy\ beam^{-1}\ km\ s^{-1}}$ for J0107a, and $\sigma=0.035\ \mathrm{Jy\ beam^{-1}\ km\ s^{-1}}$ for J0107b (see Section \ref{subsec:line_identification} and Figures \ref{fig:J0107a}, \ref{fig:J0107b}).}
}
 \label{fig:HST}
\end{figure*}
\normalsize{}


\section{Results} \label{sec:Results}


\subsection{Molecular gas mass {by emission lines}}
\label{sec:Mgas_alpha}

The molecular gas mass can be calculated from both the CO and [C\,{\sc i}] line luminosities. 
The molecular gas mass derived from the CO line luminosity has an uncertainty caused by {the excitation and} the choice of a CO-to-H$_2$ conversion factor $\alpha_{\mathrm{CO}}$. 
Although the molecular gas mass derived from [C\,{\sc i}](1--0) has a similar uncertainty due to the [C\,{\sc i}] abundance, this can be used for reasonable estimation of H$_2$ and molecular gas mass because of its simple partition function and chemistry \citep{Alaghband-Zadeh13,Saito20}.
In this section, we first calculate $M_{\rm{gas}}$ with the [C\,{\sc i}] line luminosity, 
and compare these results with $M_{\rm{gas}}$ which are derived with CO line luminosity and typical $\alpha_{\mathrm{CO}}$ for high-redshift galaxies.
Hereafter, we adopt some parameters and equations for submillimeter galaxies (SMGs) from previous studies,
{such as CO line luminosity ratio, dust temperature $T_{\rm{dust}}$, Equations (\ref{eq:Scoville}), and (\ref{eq:Hainline}),}
because the properties of J0107a and J0107b based on the analysis in this study{, such as  submillimeter fluxes and star formation rate,} are  SMG-like  {(see Table \ref{tab:continuum} and Section
\ref{subsubsec:Far-IR luminosity and SFR} about J0107a).}

We calculate the
$\rm H_2$ mass
using the formula given by \cite{Alaghband-Zadeh13} (and references therein):

\begin{eqnarray}
M_{\mathrm{H_{2}}}=1375.8\ \frac{D_L^2}{1+z}\left(\frac{X_{\mathrm{C_I}}}{10^{-5}}\right)^{-1}\left(\frac{A_{10}}{10^{-7}\  \mathrm{s^{-1}}}\right)^{-1} \nonumber  \\
\times Q_{10}^{-1}\frac{S_{\mathrm{C_I}}}{\mathrm{Jy\ km\ s^{-1}}}\ \ [M_{\odot}], 
\label{eq:a}
\end{eqnarray}

where $D_L$ is the luminosity distance in Mpc {(20656 Mpc for J0107a and 19073 Mpc for J0107b)}, $S_{\rm{C_I}}$ is the flux density of the [C\,{\sc i}](1--0) line, $A_{10}=7.93\times10^{-8}\ \mathrm{s^{-1}}$ is the Einstein A coefficient, {$X_{\mathrm{C_I}} = [\mathrm{C_I}]/[\mathrm{H_2}]=(5.0-7.9)\times10^{-5}$ 
is the [C\,{\sc i}] abundance relative to $\rm H_2$ for SMGs at $z>2.5$}
{ \citep[][and the references therein]{Valentino18}}, and $Q_{10}=Q_{10}(n,T_{\rm{ex}})=0.49\pm0.02$ \citep{Alaghband-Zadeh13} is the partition function, which depends on the gas excitation conditions \citep{Papadopoulos04} {and here the excitation temperature is assumed to be about 30 K \citep{Alaghband-Zadeh13}}. 
While $A_{10}$ is naturally a physical constant, $X_{\rm{C_I}}$ and $Q_{10}$ are empirical parameters.

By multiplying the $\rm H_2$ mass, calculated from Eq. (\ref{eq:a}),
by 1.36 to account for the $\rm He$ contribution, we derive the total molecular gas mass $M_{\rm gas}$: 
\begin{eqnarray}
M_{\rm gas}({\rm J0107a}) &=& ( 11.2 \pm 3.1 ) \times 10^{10} \quad M_{\odot}, \nonumber\\
M_{\rm gas}({\rm J0107b}) &=& ( 4.2 \pm 1.2 ) \times 10^{10} \quad M_{\odot}. \nonumber
\label{eq:2}
\end{eqnarray}
The dominant factor in the error is the uncertainty in $X_{\mathrm{C_I}}$.





{On the other hand, $M_{\rm{gas}}$ can be derived from the CO(1--0) luminosity and $\alpha_{\rm{CO}}$ as: }

\begin{equation}
    \frac{M_{\rm{gas}}}{M_{\odot}}=\alpha_{\mathrm{CO}}\ \frac{L'(\mathrm{CO})_{1-0}}{\mathrm{K~km\ s^{-1}~pc^2}}.
\end{equation}

{We here estimate $M_{\rm{gas}}$ by assuming $\alpha_{\rm{CO}} = 0.8$ \citep{Downes98}, which is commonly adopted for high-redshift galaxies, and converting CO(4--3) and CO(3--2) luminosities to that of CO(1--0) based on SMGs' line ratios \citep[$L'(\mathrm{CO})_{4-3}/L'(\mathrm{CO})_{1-0}=0.32\pm0.05$, and $L'(\mathrm{CO})_{3-2}/L'(\mathrm{CO})_{1-0}=0.60\pm0.11$;][]{Birkin21}. 
{The results are $M_{\mathrm{gas}}\,(\mathrm{J0107a})=(13.4\pm1.7)\times10^{10}\ M_{\odot}$ and 
$M_{\mathrm{gas}}\,(\mathrm{J0107b})=(5.0\pm0.7)\times10^{10}\ M_{\odot}$ for CO(4--3) luminosities, and $M_{\mathrm{gas}}(\mathrm{J0107a})= (9.2\pm1.9)\times10^{10}\ M_{\odot}$ and $M_{\rm{gas}}(\mathrm{J0107b})=(3.6\pm0.8)\times10^{10}\ M_{\odot}$ for CO(3--2) luminosities.}
These results are consistent with the $M_{\rm{gas}}$ based on [C\,{\sc i}] within uncertainties.}

\begin{deluxetable*}{cccccc}
\tablecaption{Continuum fluxes for the targets.}
\tablewidth{0pt}
\tablehead{
\colhead{Facility}&\colhead{Center}&\colhead{beam size}
&\colhead{$S_{\rm{J0107a}}$  [mJy]}&\colhead{$S_{\rm{J0107b}}$  [mJy]}&\colhead{rms [mJy beam$^{-1}$]}}
\startdata
 ALMA/Band 3&3.0\,mm&$1\farcs14\times1\farcs07$&$<0.27$&$<0.27$&0.09\\
 ALMA/Band 4&2.2\,mm&$1\farcs16\times0\farcs95$&$0.61\pm0.12$&$<0.42$&0.14\\
ALMA/Band 6&1.3\,mm&$0\farcs57\times0\farcs50$&$3.4\pm0.3$&$0.96\pm0.2$&0.07\\
ALMA/Band 7&887\,$\mu$m&$0\farcs18\times0\farcs15$&$7.9\pm1$&$5.1\pm1.2$&0.15\\
\hline
SMA&1.3\,mm&&$5.2\pm1.3$&&1.21\\
\enddata
\label{tab:continuum}
\tablecomments{Upper limits are 3 $\sigma$. The SMA data in the last row are from \cite{Tamura14}. }
\end{deluxetable*}

\subsection{Molecular gas mass {by dust continuum}}
 \label{subsubsec:calculation of stellar mass and interstellar mass}
 
{For local galaxies, ULIRGs, and redshifted SMGs observed at wavelengths longer than 250 $\mu$m, in which we may  assume that the emission is in the Rayleigh–Jeans tail, 
\cite{Scoville16} presented that $M_{\rm{gas}}$ can be derived from the dust continuum emission as follows:}

\begin{eqnarray}
\label{eq:Scoville}
\frac{M_{\rm{gas}}}{10^{10}M_{\odot}}=1.78\,(1+z)^{-4.8}\frac{S_{\nu_{\rm{obs}}}}{\mathrm{mJy}}\left(\frac{\nu_{\rm{obs}}}{353\ \mathrm{GHz}}\right)^{-3.8} \nonumber \\
\times\left(\frac{6.7\times10^{19}}{\alpha_{\mathrm{353\,GHz}}}\right)\frac{\Gamma_0}{\Gamma_{\rm{RJ}}}\left(\frac{D_L}{\mathrm{Gpc}}\right)^2,
\end{eqnarray}
where
\begin{equation}
\frac{\alpha_{\mathrm{353\,GHz}}}{\mathrm{erg~s^{-1}~Hz^{-1}}~M_{\odot}^{-1}}=L_{\mathrm{353\,GHz}}/M_{\mathrm{gas}}\ ,
\end{equation}
\begin{equation}
\Gamma_{\rm{RJ}}(T_{\rm{dust}},\nu_{\rm{obs}},z)=\frac{h\nu_{\rm{obs}}(1+z)/kT_{\rm{dust}}}{e^{h\nu_{\rm{obs}}(1+z)/kT_{\rm{dust}}}-1}\ ,
\end{equation}
and $\Gamma_0=\Gamma_{\rm{RJ}}(T_{\rm{dust}},353\ \mathrm{GHz},0)$.

{We use our ALMA/band 6 data for calculation of $M_{\rm{gas}}$ of J0107a.} 
Because our current data set cannot constrain $T_{\rm{dust}}$, we assume $T_{\rm{dust}}$ = 40 K which is given by the surveys of SMGs and galaxies in $0<z<4$  \citep{Cunha15,Schreiber18}. {Here, we should note that this $T_{\rm{dust}}$ is luminosity-weighted and is likely higher than mass-weighted $T_{\rm{dust}}$, such as that suggested in \cite{Scoville14}.} 
Consequently, $M_{\rm{gas}}$ of J0107a is calculated as below:
\[M_{\rm{gas}}\,(\mathrm{J0107a})=(3.2\pm1.6)\times10^{11}\ M_{\odot}\ .\]
{Here, we adopt $\alpha_{\mathrm{353\,GHz}}=8.4\times10^{19}\  \mathrm{erg~s^{-1}~Hz^{-1}}~M_{\odot}^{-1}$, which is derived as the mean value for $z\sim2$ SMGs in \cite{Scoville16}. We also set a typical uncertainty of the derived $M_{\mathrm{gas}}$ as $\pm50\%$  \citep{Scoville14}.}
If $T_{\rm{dust}}=30$ K and 50 K are adopted, $M_{\rm{gas}}$ becomes $M_{\rm{gas}}(\mathrm{J0107a})=(3.6\pm1.8)\times10^{11}\ M_{\odot}$ and $M_{\rm{gas}}\,(\mathrm{J0107a})=(3.0\pm1.5)\times10^{11}\ M_{\odot}$, respectively.
{Hence, the result may change from approximately 6\% to 12\% with a $T_{\rm{dust}}$ error of 10 K.}
{We calculate the $M_{\rm{gas}}$ of J0107b in the same manner as follows:}
\[M_{\rm{gas}}\,(\mathrm{J0107b})=(9.5\pm4.8)\times10^{10}\ M_{\odot}\ .\]
{The result of J0107b also changes approximately 6\% to 12\% with a $T_{\rm{dust}}$ error of 10 K. }

{The similarity between the cold gas mass derived from the lines and from the dust continuum supports the derived cold gas mass.}

\subsection{Far-IR luminosity and SFR}
\label{subsubsec:Far-IR luminosity and SFR}

In star-forming galaxies (SFGs), most infrared emissions are believed to be emitted from warm dust around young stars, and, thus, their SFR is calculated from the infrared luminosity $L_{\rm IR}$ \citep{Kennicutt98,Carilli13}.
Although we can also assume AGN to be a heat source, 
\cite{Brown19} reported that the AGN contribution to $L_{\rm{IR}}$ is typically smaller than $10$\% when $L_{\rm{IR}}\sim10^{13}\ L_{\odot}$.
Here, we estimate the SFR of J0107a and J0107b using the far-infrared luminosity $L_{\rm{FIR}}$ because of the lack of $L_{\rm{IR}}$ data. 
By assuming the gray body model $F_{\nu}\propto \nu^{\beta}B(\nu, T)$, {we derive $L_{\rm{FIR}}$ with the formula in \cite{De-Breuck03}:}
\begin{eqnarray}
L_{\rm{FIR}}=4\pi\Gamma[\beta+4]\zeta[\beta+4]D_L^2\left(\frac{h\nu_{\rm{rest}}}{kT_{\rm{dust}}}\right)^{-(\beta+4)}\nonumber \\
\times\left(e^{h\nu_{\rm{rest}}/kT_{\rm{dust}}}-1\right)S_{\rm{obs}}\nu_{\rm{obs}}\ ,
\end{eqnarray}

where $\beta$ is the beta index, $\Gamma[x]$ is the Gamma function, and $\zeta[x]$ is the zeta function.
{Assuming $\beta=1.5$ and $T_{\rm{dust}}=40$ K, we
calculate $L_{\rm{FIR}}$ of  J0107a and J0107b using this formula as}
\[L_{\rm{FIR}}(\mathrm{J0107a})=(10.3\pm0.8) \times10^{12}\ \ L_{\odot}\ ,\]
\[L_{\rm{FIR}}(\mathrm{J0107b})=(3.0\pm0.6)\times10^{12} \ \ L_{\odot}\ .\]

{Here, we used band 6 data, in which both targets are clearly identified. 
For a sanity check, we also calculate $L_{\mathrm{FIR}}$ with other continuum data.
For J0107a, we derived $L_{\mathrm{FIR}}<12.3\times10^{12}\ L_{\odot}$, $ L_{\mathrm{FIR}}=(10.2\pm2.0)\times10^{12}\ L_{\odot}$, and $L_{\mathrm{FIR}}=(9.0\pm1.1)\times10^{12}\ L_{\odot}$ with band 3, 4, and 7 data, respectively. 
These results are all consistent with each other. We calculate for J0107b with the same manner as  $L_{\mathrm{FIR}}<12.3\times10^{12}\ L_{\odot}$, $ L_{\mathrm{FIR}}<7.2\times10^{12}\ L_{\odot}$, and $L_{\mathrm{FIR}}=(5.8\pm1.4)\times10^{12}\ L_{\odot}$ with band 3, 4, and 7 data, respectively. They are also comparable to each other. $L_{\mathrm{FIR}}$ of J0107b derived from band 7 data is a little larger than that from band 6 data. This may be because of poorer fitting in band 7 data due to apparent multiple peaks. } 

 Regarding $L_{\rm{FIR}}$, the results change by a factor of 2 when $T_{\rm{dust}}$ changes by 10 K. 
The results also depend on $\beta$. \cite{Chapin09} derived $\beta=1.75$ for 29 SMGs with a median redshift $z=2.7$ with a 1.1 mm survey. The \cite{Planck11b} similarly suggested $\beta=1.8\pm0.1$ based on the all-sky observation results, and \cite{Scoville14} adopted $\beta=1.8$ after this result. If we adopt $\beta=1.8$ for our calculation of $L_{\rm{FIR}}$, the results with $T_{\rm{dust}}=40$ K increase by about a factor of 1.5 {($L_{\mathrm{FIR}}(\mathrm{J0107a})=1.7\times10^{13}\ L_{\odot}$ and $L_{\mathrm{FIR}}(\mathrm{J0107b})=5.0\times10^{12}\ L_{\odot}$, respectively). Consequently, the total systematic uncertainty of $L_{\mathrm{FIR}}$, due to the selection of $T_{\mathrm{dust}}$ and $\beta$, is about a factor of 3.}
 
The relationship between SFR and $L_{\rm{FIR}}$ is shown {in \cite{Genzel10} }as follows:
\begin{equation}
\label{eq7}
\log_{10} \left(\frac{SFR}{M_{\odot}\,\mathrm{yr^{-1}}}\right)=\log_{10} \left(\frac{L_{\rm{FIR}}}{L_{\odot}}\right)+\log_{10}(1.3)-10, 
\end{equation}
where $\log(1.3)$ is a correction factor between $L_{\rm IR}$ and $L_{\rm FIR}$ \citep{Garicia08}, {and the typical uncertainty of this equation is about $\pm50\%$ \citep{Genzel10}.}
The derived SFRs are $1.3 \times 10^3$ $M_{\odot}$~yr$^{-1}$ and $3.9 \times 10^2$ $M_{\odot}$~yr$^{-1}$ for J0107a and J0107b, respectively, and the typical systematic uncertainties of these SFRs are {about a factor of 4.5, which is mainly attributed to the uncertainties of $L_{\mathrm{FIR}}$ and the Equation (\ref{eq7}).}
The physical quantities derived from our data analysis are summarized in Table \ref{tab:quantities} and discussed in Section \ref{sec:Discussion}.

\tabletypesize{\scriptsize}
\begin{deluxetable*}{cccccccc}
\tablecaption{Physical quantities derived by data analysis.}
\tablewidth{0pt}
\tablehead{
\colhead{Target}&\colhead{$z$}&\colhead{$M_{\rm{gas}}$([C\,{\sc i}]) [$M_{\odot}$]}&\colhead{{$M_{\rm{gas}}$(CO) [$M_{\odot}$]}}&\colhead{{$M_{\rm{gas}}(\mathrm{dust})$} [$M_{\odot}$]}&\colhead{$L_{\rm{FIR}}$ [$L_{\odot}$]}&\colhead{SFR [$M_{\odot}$ yr$^{-1}$]}
}
\startdata
J0107a&$2.4666\pm0.0002$&$(11.2\pm3.1)\times10^{10}$&$(13.4\pm1.7)\times10^{10}$&$(32\pm16)\times10^{10}$&$10.3\times10^{12}$&$1.3\times10^3$\\
J0107b&$2.3100\pm0.0002$&$(4.2\pm1.2)\times10^{10}$&$(5.0\pm0.7)\times10^{10}$&$(9.5\pm4.8)\times10^{10}$&$3.0\times10^{12}$&$3.9\times10^2$\\
\enddata
\tablecomments{$z$ : spectroscopic redshift; $M_{\rm{gas}}$([C\,{\sc i}]) : molecular gas mass derived by [C\,{\sc i}](1--0) line intensity; {$M_{\rm{gas}}$(CO) : molecular gas mass derived by CO(4--3) line luminosity with $\alpha_{\mathrm{CO}}=0.8$}; {$M_{\rm{gas}}(\mathrm{dust})$ : molecular gas mass derived by dust continuum emission;} $L_{\rm{FIR}}$ : far-infrared luminosity derived by dust continuum emission; SFR : star formation rate derived by $L_{\rm{FIR}}$. 
{SFR have typical uncertainties of {about a factor of 4.5}.}
}
\end{deluxetable*}
\label{tab:quantities}

\section{Kinematic modeling} \label{sec:kinematic modeling}
In this section, we perform kinematic modeling of a rotating disk to derive a rotation curve and estimate some dynamical properties.
Here, we discuss only J0107a.
Although the CO(4--3) line of J0107b indicates the rotational motion (see Figure \ref{fig:J0107b}), the S/N is not enough for the kinematic modeling.

\subsection{Method and results} \label{subsec:method and results}

We model the disk rotation of J0107a using the Markov chain Monte
Carlo (MCMC) 
method with GalPaK$^{\rm 3D}$ \citep{Bouche15}. 
We use the CO(4--3) data cube (2D image and frequency dimension) for the modeling, which has a better S/N than those of CO(3--2) or [C\,{\sc i}](1--0). 
The algorithm directly compares the data cube with a disk parametric model with ten free parameters:
coordinates of the galaxy center ($x_c, y_c, z_c$), flux, half-light radius, inclination angle ($i$), position angle (PA), turnover radius of the rotation curve, deprojected maximum rotation velocity, and intrinsic velocity dispersion.
We first perform modeling with ten free parameters, and then repeat it by setting the initial values for $x_c$, $y_c$, $z_c$, and $i$. 
We assume a rotational velocity with a hyperbolic tanh profile in this model.
The uncertainty is the 95\% confidence interval (CI) calculated from the last 60\% of the MCMC chain for 20,000 iterations.
{We here set the random scale of MCMC chain as 0.4 to obtain its acceptance rate of 30--50\%, which is suggested by GalPaK$^{\rm{3D}}$. By doing this, we achieved the acceptance rate of $\sim38\%$ in this modeling.
Furthermore, by setting a beam size, the effect of beam smearing is considered  in the modeling. }

\begin{figure}[h]
\centering
\includegraphics[bb=0 0 640 480,width=11cm]{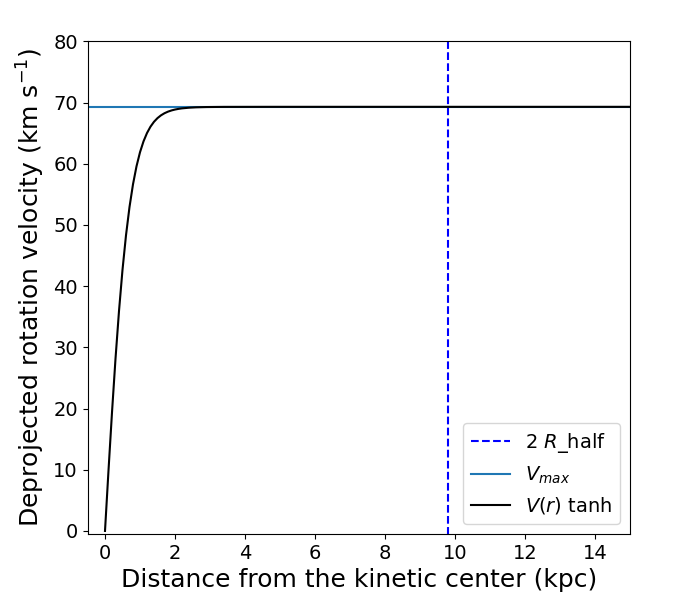}
\vspace{0.2cm}
\caption{
The model rotation curve of J0107a calculated from GalPaK$^{\rm{3D}}$ (black solid curve). The horizontal blue solid line indicates the calculated maximum velocity of rotation and the vertical blue dashed line indicates the twice of half-light radius. 
}
\label{fig:rot_curve}
\end{figure}

\begin{figure*}      
\gridline{\fig{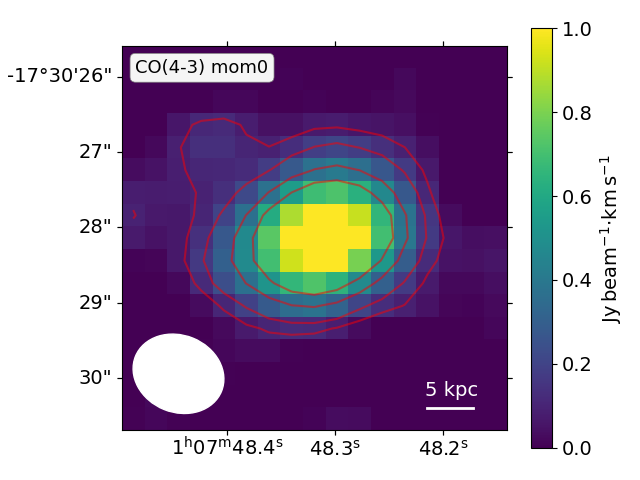}{0.32\textwidth}{(a)}
          \fig{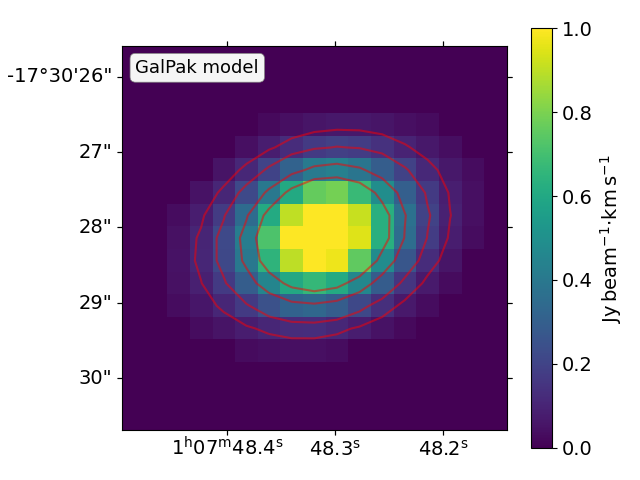}{0.32\textwidth}{(b)}
          \fig{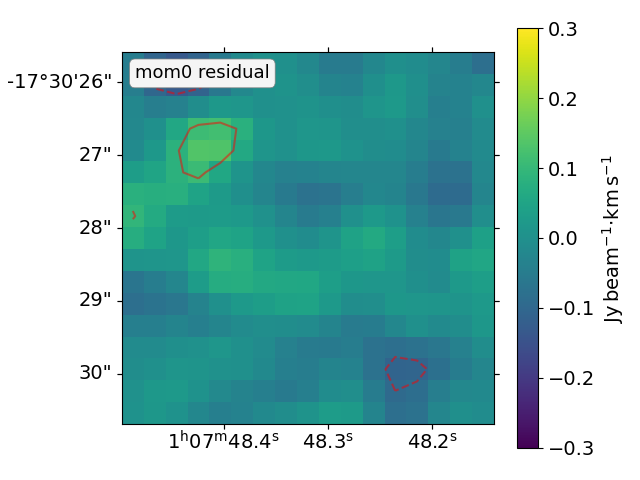}{0.32\textwidth}{ (c)}
          }
\gridline{\fig{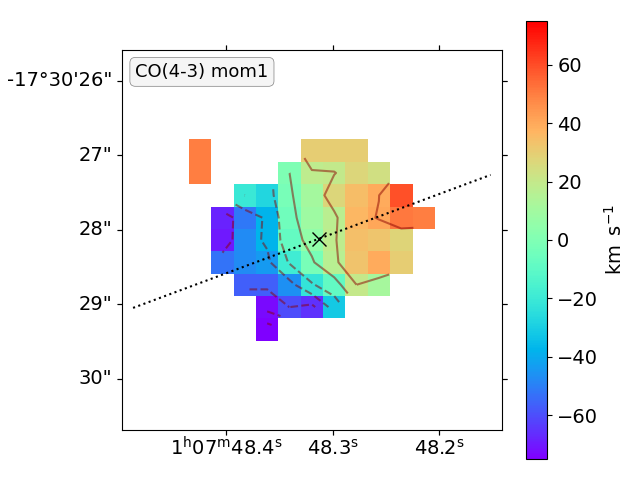}{0.32\textwidth}{(d)}
          \fig{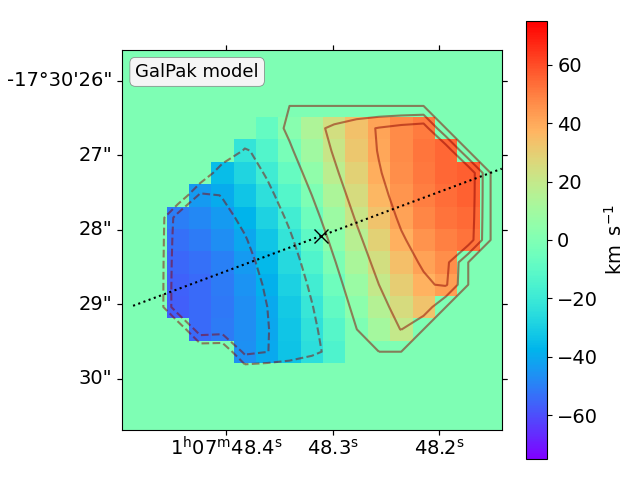}{0.32\textwidth}{(e)}
          \fig{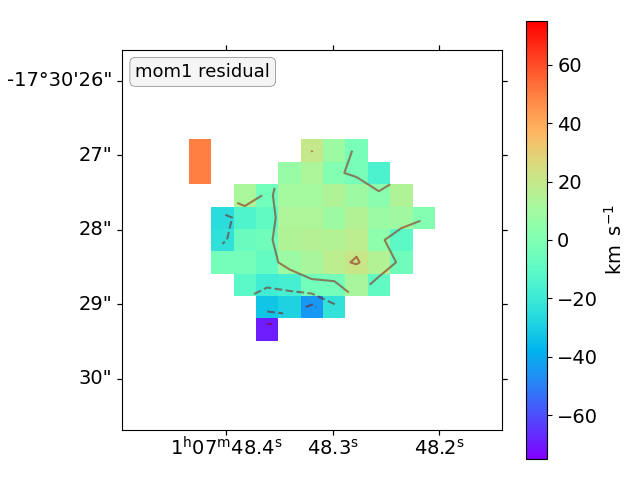}{0.32\textwidth}{ (f)}
          }
          \caption{Comparison of the observed (left panels) and modeled (middle panels) CO(4-3) data of J0107a, along with their residuals (right panels). (a): The observed CO(4--3) integrated intensity map. The contours show 5, 10, 20, and 30 $\sigma$ of the data, and contours with the same levels are also shown in (b). (b): The modeled CO(4--3) integrated intensity map from the best-fit GalPaK$^{\rm{3D}}$ results. (c): The residual of the CO(4-3) integrated intensity map, (a)-(b). The contours show $\pm5\sigma$ of the data. 
          (d): The observed CO(4-3) velocity map. The contour interval in (d), (e), and (f) is 20 km s$^{-1}$, and negative contours are dashed. (e): The modeled CO(4-3) velocity map from the GalPaK$^{\rm{3D}}$ model. {The velocities in the position-velocity diagram in  Figure \ref{fig:pv_diagram} are extracted along the dotted line shown in (d) and (e).} (f): The residual of the CO(4-3) velocity map, (d)-(e).} 
\label{fig:galpak_compare}
\end{figure*}

\begin{figure}[h]
\centering
\includegraphics[bb=0 0 640 480,width=12cm]{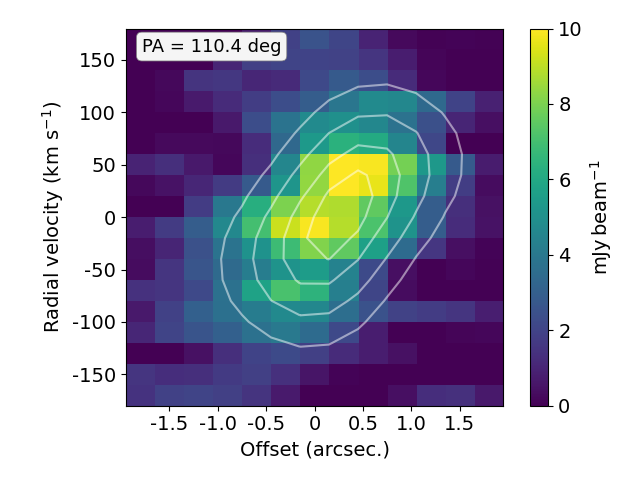}
\vspace{0.2cm}
\caption{
The position-velocity diagram of J0107a along the major axis with PA = $110.4\tcdegree$ (the dashed lines in Figure \ref{fig:galpak_compare} (d, e)). The background color is from the ALMA CO(4-–3) data while the contours are from GalPaK$^{\rm{3D}}$ results. Contour levels are 2, 3.5, 5, and 6 mJy beam$^{-1}$.
}
\label{fig:pv_diagram}
\end{figure}

The initial value for $i$ is set to be $60\tcdegree$ based on the major to minor axis ratio of the beam-deconvolved source size of the 1.3 mm continuum. 
{We confirm that setting initial values to be $20\tcdegree$ or $40\tcdegree$ does not largely change the results, and hereafter present results with $i = 60\tcdegree$ only.}
We then obtain a GalPaK$^{\rm{3D}}$ model,
and the inclination angle converges to approximately $63\tcdegree$. 
This result indicates that J0107a is unlikely to be face-on, and the major/minor ratio of the model ($=2.2\pm0.1$) is consistent with the observed source sizes of 1.3 mm, 0.89 mm, and CO(4-3) emission.
In addition, the half-light radius of the model is consistent with the deconvolved {half width at half maximum} of J0107a ($=5.4\pm0.5$ kpc), which is derived by the CASA task \sf{imfit}\rm{} for the CO(4--3) source in the band 4 data.
We also obtain the maximum rotation velocity as $v_{\rm{max}}=69.3\pm4.8$ km s$^{-1}$, and the velocity dispersion as $\sigma_v=54.6\pm1.8$ km s$^{-1}$. 
{In GalPaK$^{\mathrm{3D}}$, velocity dispersion is estimated by assuming three components, one of which is an intrinsic velocity dispersion \citep{Bouche15}, and $\sigma_v$ of J0107a here is the result for the intrinsic velocity dispersion. Therefore, we hereafter treat $\sigma_v$ of J0107a as an intrinsic value.}
We summarize the results of the GalPaK$^{\rm{3D}}$ model in Table \ref{tab:GalPak}. 
Figure \ref{fig:rot_curve} shows the output rotation curve, and
Figure \ref{fig:galpak_compare} shows a comparison of the GalPaK$^{\rm{3D}}$ model with the observational data of J0107a on CO(4--3) intensity and velocity maps. 
{Figure \ref{fig:pv_diagram} shows the position-velocity diagram of J0107a and the contours of its model by GalPaK$^{\rm{3D}}$. 
Both of the velocities in this diagram are extracted along the lines shown in the black dotted lines in Figure \ref{fig:galpak_compare}(d) and (e), whose PA is $110.4\tcdegree$. 
As we can see in Figure \ref{fig:pv_diagram}, the velocity of J0107a changes continuously from $v\sim-120\ \mathrm{km\ s^{-1}}$ to $\sim+120\ \mathrm{km\ s^{-1}}$, and there are apparently two peaks at $v\sim-20\ \mathrm{km\ s^{-1}}$ and $v\sim+40\ \mathrm{km\ s^{-1}}$ in this diagram. 
This result suggests that J0107a is less likely to be a galaxy merger. All of these results suggest that J0107a has a rotating disk.  }
Additionally, we show the plot of cross-correlations in the Markov chain in Appendix Figure \ref{fig:correlation}.


\scriptsize
\begin{table}[h]
\begin{center}
\caption{
The results of the GalPaK$^{\rm{3D}}$ kinematic modeling of J0107a. 
}
\label{tab:GalPak}
\begin{tabular}{ccc}
\hline \hline
J0107a&value&95\% CI\\ \hline
flux [Jy beam$^{-1}$]&$1.83\pm0.02$&$[1.78, 1.88]$\\
half-light radius [kpc]&$4.91\pm0.1$&$[4.71, 5.13]$ \\
inclination [deg]&$63.2\pm1.7$&$[59.8, 66.8]$\\
pa [deg]&$290.4\pm1.5$&$[287.5, 293.2]$ \\
turnover radius [kpc]&$0.697\pm0.772$&$[0.05, 2.71]$ \\
maximum velocity [km s$^{-1}$]&$69.31\pm4.79$&$[64.45, 82.53]$ \\
velocity dispersion [km s$^{-1}$]&$54.57\pm1.78$&$[51.48, 58.35]$ \\ \hline
\end{tabular}
\vspace{0.3cm}
\end{center}
\end{table}

\subsection{Dynamical mass}\label{subsec:dynamical mass} 

\normalsize{}
{In the calculation of a dynamical mass, $M_{\rm{dyn}}$, we use the results of GalPaK$^{\rm{3D}}$. 
As for the radius, we regard the twice of half-light radius, which corresponds to $r=9.8\pm0.2$ kpc ($=1.2\pm0.02$ arcsec.), as the radius of J0107a. }
{
We use this $r$ and the best-fit value of $v_{\mathrm{max}}$ from the model to estimate
$M_{\rm{dyn}}$, and derive it} as follows:

\[M_{\rm{dyn}}\,(\mathrm{J0107a})=\frac{rv_{\rm{max}}^2}{G}=(1.1\pm0.2) \times10^{10} \ \ M_{\odot}\ .\]

Here, $G$ is the gravitational constant, and $1.1\times10^{10}\ M_{\odot}$ is the 50th percentile of the results of $M_{\rm{dyn}}$ for each {12,000} iterations after MCMC burn-in.   
This is an order of magnitude smaller than $M_{\mathrm{gas}}(\mathrm{dust})$ derived in Section~\ref{subsubsec:calculation of stellar mass and interstellar mass}. The possible origins of this discrepancy are discussed in the next section. 

\section{Discussion} \label{sec:Discussion}

Here, we {mainly} discuss the physical properties of J0107a. Specifically, we focus on the ratio of maximum rotation velocity $v_{\rm{max}}$ to velocity dispersion $\sigma_v$, $v_{\rm{max}}/\sigma_v$, to assess the dynamic hotness of the gas disk, and the molecular gas fraction $f_{\rm{gas}} =M_{\rm{gas}}/(M_{\rm{gas}}+M_{\star})$ , where $M_{\star}$ is the stellar mass of the galaxy, to characterize the evolutionary state of the system. 
The possible cause of the discrepancy between $M_{\mathrm{gas}}$ and $M_{\mathrm{dyn}}$, which is found to be of an order of magnitude, is also discussed in this section. 





\subsection{Dynamical properties of the gas disk}\label{subsec:characteristics of the galactic disk}


We obtain $v_{\rm{max}}/\sigma_v = 1.3 \pm 0.1$ using the best-fit kinetic parameters in Section \ref{sec:kinematic modeling}.
Here, 1.3 is the 50th percentile of the results for each {12,000} iterations after MCMC burn-in, similar to $M_{\rm{dyn}}$. 
{The error of $v_{\rm{max}}/\sigma_v$ is also calculated with the use of each result of 12,000 iterations. 
As a more conservative estimate on the error in $\sigma_v$, we adopt the error of the spectral line width derived by the Gaussian fitting (Table \ref{tab:lineresults}). The largest error is 34 km s$^{-1}$ (in FWHM) for the [C\,{\sc i}] line, which corresponds to 13 km s$^{-1}$ in standard deviation $\sigma$ when corrected for the inclination angle.
Consequently the error of $v_{\rm{max}}/\sigma_v$ can be estimated as 0.3, which is still very small. We hereafter adopt this error for conservative discussion.}
This result suggests that $\sigma_v$ of  J0107a is compareble to $v_{\rm{max}}$, 
{and the disk is rather turbulent.}

\begin{figure*}      
\gridline{\fig{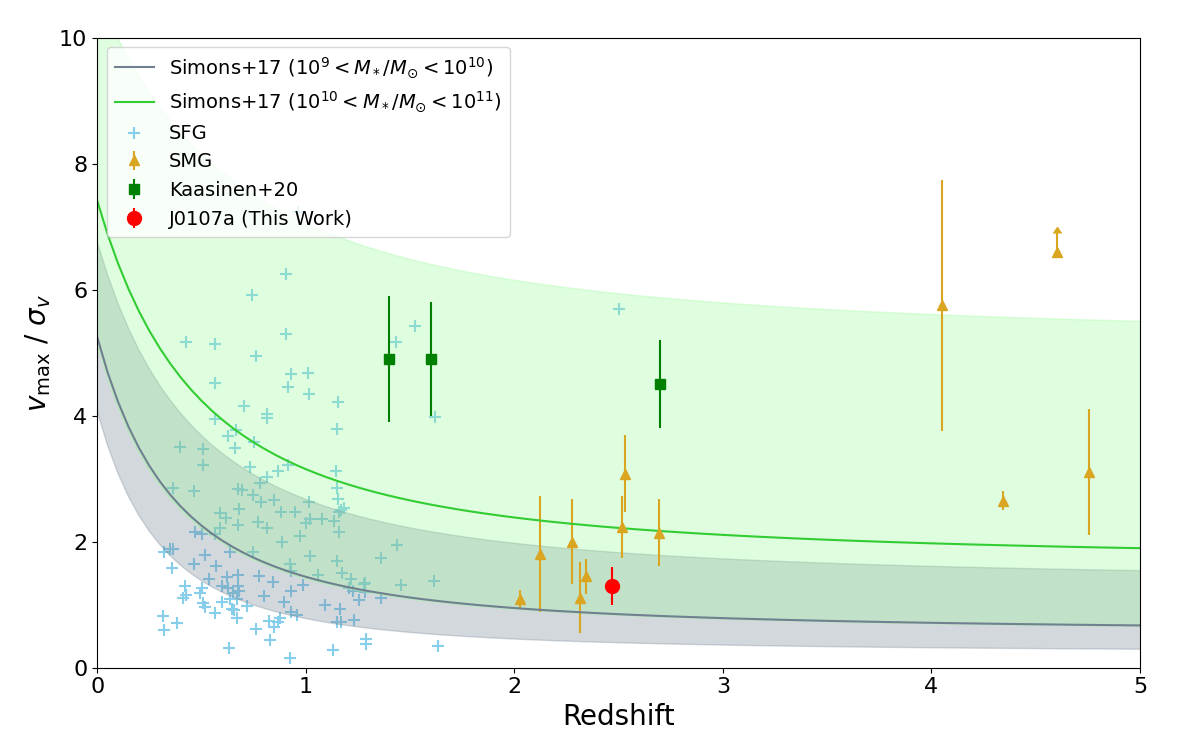}{1.0\textwidth}{}
          }
          \caption{Ratio of rotational velocity to velocity dispersion ($v_{\rm{max}}/\sigma_v$) versus redshift. The sky-blue pluses indicate SFGs  \citep[data from][]{Swinbank17,Lee19}, {the green squares indicate millimeter-wave line emitter samples in \cite{Kaasinen20}}, the brown up triangles indicate SMGs  {\citep[{data from}][]{Alaghband-Zadeh12,Hodge12,De-Breuck14,Tadaki19,Jimenez+20}}, and the red circle indicates J0107a. 
          {In \cite{Kaasinen20}, they treated their $\sigma_v$ as upper limits, but we here do not show $v_{\mathrm{max}}/\sigma_v$ of these samples as lower limits (see details in Section \ref{subsec:characteristics of the galactic disk}).} 
          We also show the fitting results for the distribution of SFGs with low stellar masses ($10^9<M_{\star}/M_{\odot}<10^{10}$) and with high stellar masses ($10^{10}<M_{\star}/M_{\odot}<10^{11}$), which are presented by \cite{Simons17}. They are indicated with a gray line and a light-green line, respectively, and the shaded areas indicate the $1\sigma$ of these fitting lines. }
\label{fig:V_sigma}
\end{figure*}

Figure \ref{fig:V_sigma} shows the relation between $v_{\rm{max}}/\sigma_v$ of SFGs {\citep[data from][]{Swinbank17,Lee19}}, SMGs {\citep[data from][]{Alaghband-Zadeh12,Hodge12,De-Breuck14,Tadaki19,Jimenez+20}}, millimeter-wave line emitter samples in \cite{Kaasinen20}, and J0107a. 

\cite{Burkert10} mentioned that $z\sim1.5$--$3.5$ SFGs have large gas velocity dispersions of $30$--$120$ km s$^{-1}$ and ratios of $v_{\rm{max}}/\sigma_{v}\sim1$--$6$. 
This is supported by the observations of $z=0.3$--$1.7$ SFGs \citep{Swinbank17}, and SFGs in a protocluster at $z\sim2.5$ \citep{Lee19}.  
\cite{Alaghband-Zadeh12} investigated the H$\alpha$ velocity fields of SMGs in $z\sim2$--3 and derived $v_{\rm{max}}/\sigma\sim1$--3 (average is $1.9\pm0.2$). 
\cite{Kaasinen20} also reported $v_{\rm{max}}/\sigma_{v}\sim5$ for their millimeter-wave line emitter samples in $z=1.4$--$2.7$, and noted that this value was higher than those of typical high-redshift galaxies due to a selection bias. 
They also noted that the velocity dispersions of their samples were global estimates that include dispersions due to motion along the line-of-sight (i.e. due to motion inside a thick disk, or, motions due to warps), and they treated their velocity dispersions as upper limits.
{However, this discussion can be applied to the $\sigma_v$ of other samples, which were derived with, for example, line width, and does not significantly affect our results. We therefore do not strictly consider it here.}
Because of this reason, we here do not treat $\sigma_v$ of Kaasinen’s samples as upper limits, as with other samples.

Consequently, from a quantitative point of view, the $v_{\rm{max}}/\sigma_v$ ratio of J0107a seems smaller than those of other line emitters, and at the same level as those of  SMGs at similar redshifts.  

\cite{Simons17} presented that the $v_{\rm{max}}/\sigma_v$ ratio of an SFG is affected by stellar mass; hence, we show the fitting results for the distribution of $v_{\rm{max}}/\sigma_v$ ratios of typical SFGs with low stellar mass ($M_{\star}\sim10^9-10^{10}\ M_{\odot}$), and high stellar mass ($M_{\star}\sim10^{10}-10^{11}\ M_{\odot}$), respectively, in Figure \ref{fig:V_sigma}.
In this figure, the $v_{\rm{max}}/\sigma_v$ ratios of almost all line emitters and SMGs, which have stellar masses of approximately $10^{11}\ M_{\odot}$, are consistent with the fitting result for high-$M_{\star}$ SFGs within uncertainties. 
The SFG samples in this figure have a very wide range of stellar mass ($M_{\star}\sim10^6-10^{11}\ M_{\odot}$), but their $v_{\rm{max}}/\sigma_v$ are also consistent with the fitting lines within uncertainties.

The $v_{\rm{max}}/\sigma_v$ ratio of J0107a is consistent with the fitting for both low and high stellar masses within the error. The stellar mass of J0107a is derived as $M_{\star}<1.0\times10^{11}\ M_{\odot}$ (see Section \ref{subsec:gas_fraction}), but considering the gravitational lensing (see Section \ref{subsec:galactic mass and star formation}), {it may be more appropriate to conclude that the $v_{\rm{max}}/\sigma_v$ ratio of J0107a is consistent with the fitting for low stellar mass, whereas it is difficult to constrain the range of $M_{\star}$ by the $v_{\rm{max}}/\sigma_v$ ratio alone.} \\


\subsection{Molecular gas fraction}\label{subsec:gas_fraction}

To obtain $f_{\rm{gas}}$, we at first calculate $M_{\star}$.
\cite{Hainline11} used a sample of $\sim70$ SMGs to derive the ratio between $M_{\star}$ and the rest-frame near-infrared ($H$-band) luminosity $L_{H}$ for two star formation histories, instantaneous starburst (IB) and constant star formation (CSF), by taking the average of best-fit model ages over the sample. 
Consequently, they obtained the relation between $M_{\star}$ and $L_{H}$ of SMGs for the population synthesis model of \cite{BC03} and CSF history as below within a factor of 2--3:
\begin{equation}
\label{eq:Hainline}
\frac{M_{\star}}{M_{\odot}}=\frac{1}{5.8}\left(\frac{L_H}{L_{\odot}}\right)\ .
\end{equation}

Considering $\lambda_H=1.65\,\mu$m, the observed wavelength is $5.72\,\mu$m with $z=z(\mathrm{J0107a})$.
Therefore, we first use the intensity data in the nearest band, IRAC/$5.8\,\mu$m data \citep{Tamura14}, which is equal to $S_{\rm{obs}}<0.1 $\,mJy (upper limit), and obtain the $L_H$ of  J0107a as follows:

\[L_H\,\sim\,4\pi D_L^2\times S_{\rm{obs}}\ d\nu<5.8\times10^{11}\ \ L_{\odot}\ ,\]

where $d\nu=4.4\times10^{13}\,$Hz is the band width of the IRAC/5.8 $\mu$m band, which is calculated by the band width in the wavelength scale $d\lambda=0.41~\mu$m in $z=z(\mathrm{J0107a})$.

Consequently, $M_{\star}$ is calculated as follows:

\[M_{\star}\,(\mathrm{J0107a})<1.0\times10^{11}\ \ M_{\odot}\ .\]

{We can also estimate $M_{\star}$ with the use of IRAC/4.5 $\mu$m and IRAC/3.6 $\mu$m data \citep{Tamura14}, as $M_{\star}=7.1\times10^{10}\ M_{\odot}$ and $5.5\times10^{10}\ M_{\odot}$, respectively, 
but we adopt the result from IRAC/5.8 $\mu$m in order to avoid possible extinction effect.
We note again that the photometry of these data is
uncertain due to the foreground emission from VV114, and we should regard the $M_{\star}$ as an upper limit. }

The $f_{\rm{gas}}$  of  J0107a is then calculated using the results of previous calculations as 
\[f_{\rm{gas}}\,(\mathrm{J0107a})>0.5\ .\]

{This fraction may contain additional uncertainty caused by differential magnification \citep{Hezaveh12,Serjeant12} between stellar emission and gas emission, considering the possibility of strong magnification of J0107a (see Section \ref{subsec:galactic mass and star formation}).}

\subsection{Why does the {molecular gas} mass exceed the dynamical mass in J0107a? }\label{subsec:galactic mass and star formation}

Considering the dark matter contribution, $M_{\rm{dyn}}$ should generally{ be consistent with, or larger than,  $M_{\rm{gas}}+M_{\star}$ within uncertainties.} 
However, we find that, for J0107a, $M_{\rm{dyn}}$ is an order of magnitude smaller than $M_{\rm{gas}}$. 
This discrepancy cannot be attributed to the $M_{\rm{gas}}$ over-estimation because {all $M_{\rm{gas}}$, yielded with the data of [C\,{\sc i}], CO lines and dust continuum, are significantly larger than $M_{\rm{dyn}}$. }
Including $M_{\star}$, 
the discrepancy becomes even larger:

\[8<\frac{M_{\rm{gas}}+M_{\star}}{M_{\rm{dyn}}}<46.\]


The most plausible reason for this result is brightening due to the
gravitational lens effect. 
\cite{Harris12} reported that 11 SMGs identified by H-ATLAS at $z\sim2.1$--$3.5$ are amplified by the gravitational lens effect. \cite{Harris12} 
claimed that the lens magnification rate $\mu$ can be estimated from
the empirical relation between the CO linewidth and luminosity for unlensed systems as



\begin{equation}
\mu=3.5\times\frac{L'(\mathrm{CO})_{\mathrm{apparent}}}{\mathrm{10^{11}\ K\ km\ s^{-1}\ pc^2}}\left(\frac{400\ \mathrm{km\ s^{-1}}}{\Delta v_{\mathrm{FWHM}}}\right)^{1.7}\ ,
\label{eq:8}
\end{equation}

where $L'(\mathrm{CO})_{\mathrm{apparent}}$ is the apparent luminosity of CO(1--0) 
, and $\Delta v_{\mathrm{FWHM}}$ is the FWHM of the observed CO(1--0) spectra.
Subsequently, $\mu$ of J0107a is calculated as
\[\mu(\mathrm{J0107a})\,\sim\,21\ \ .\]

\begin{figure*}      
\gridline{\fig{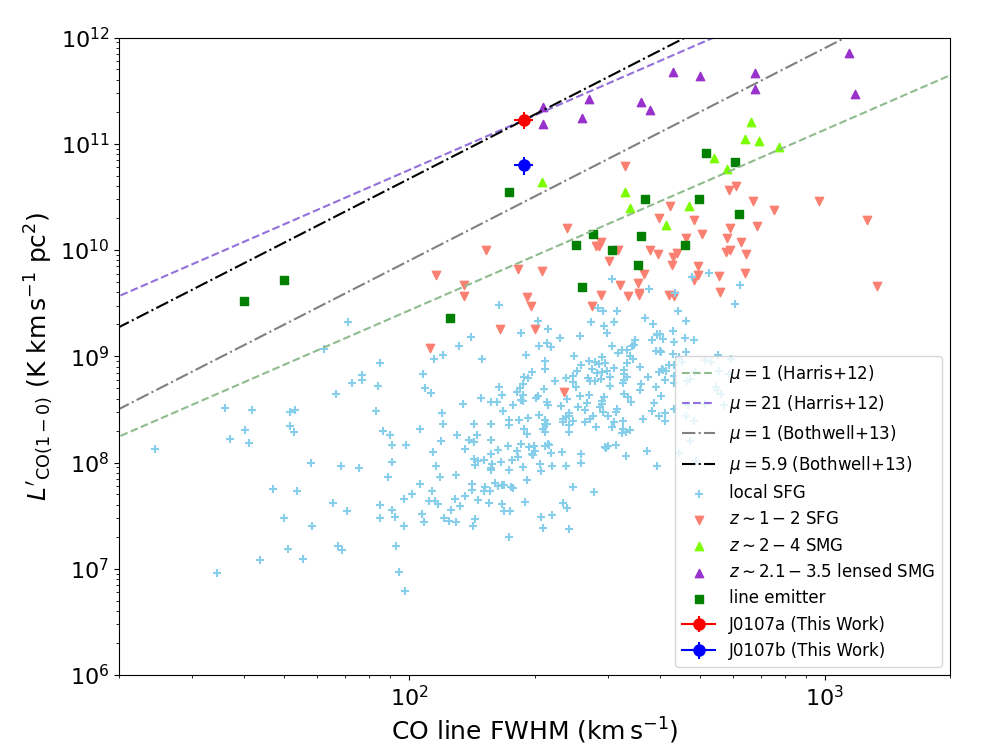}{1.0\textwidth}{}
          }
          \caption{Plot of $L'_{\mathrm{CO(1-0)}}$ versus  FWHM of CO line.
Red and blue circles indicate J0107a and J0107b, respectively. 
The data of local SFGs, which are indicated with sky-blue pluses, are taken from \cite{Bothwell14} and \cite{Saintonge17}. 
The data of $z\sim1-2$ SFGs, which are indicated with pink down triangles, are taken from \cite{Daddi10}, \cite{Magnelli12}, \cite{Magdis12}, and \cite{Tacconi13}. 
The data of $z\sim2-4$ SMGs and $z\sim2.1-3.5$ SMGs, which are indicated with green and purple up triangles, respectively, are taken from \cite{Harris12} and citation therein. 
{The data of millimeter-wave line emitters, which are indicated with green squares, are taken from \cite{Aravena19} and \cite{Kaasinen20}, while the sources in \cite{Kaasinen20} are all contained in the samples of \cite{Aravena19}, and we use the updated values from \cite{Kaasinen20} for some of these objects.}
The $z\sim2.1-3.5$ SMGs are thought to be magnified with gravitational lens effect \citep{Harris12}. 
The linewidths of all SMGs in \cite{Harris12} is that of CO(1--0), whereas most of all linewidths of local SFGs and $z\sim1-2$ SFGs in this figure are that of CO(2--1) or CO(3--2). 
We also show a power-law fit for intrinsic line luminosity versus linewidth with light-green dashed line \citep{Harris12} and gray dot dashed line \citep{Bothwell13}, and for magnified line luminosity with purple dashed line  \citep[$\mu=21$ from][]{Harris12} and black dot dashed line \citep[$\mu=5.9$ from][]{Bothwell13}.}
\label{fig:harris}
\end{figure*}

{Here we assume the linewidth ratio, $\Delta v_{\rm{CO(1-0)}}/\Delta v_{\rm{CO(3-2)}}=1.15\pm0.06$ \citep{Ivison11}, and derive $\Delta v_{\rm{FWHM}}(\rm{J0107a})\sim189$ km s$^{-1}$, while $L'(\rm{CO})_{\rm{apparent}}$ is calculated with CO(4--3) which has the best S/N in our dataset, and the SMG line ratio \citep[$L'(\mathrm{CO})_{4-3}/L'(\mathrm{CO})_{1-0}=0.32\pm0.05$;][]{Birkin21}.} 
In the same manner, $\mu(\rm{J0107b})$ is calculated to be $\mu(\rm{J0107b})\sim7.9$. 
Figure \ref{fig:harris} shows the relationship between $L'(\rm{CO})$ and $\Delta v_{\rm{FWHM}}$ for local SFGs \citep{Bothwell14,Saintonge17}, high-$z$ SFGs  \citep{Daddi10,Magnelli12,Magdis12,Tacconi13}, high-$z$ SMGs \citep{Harris12}, millimeter-wave line emitters \citep{Aravena19,Kaasinen20}, and our targets. The positions of J0107a and J0107b are higher than the average for SFGs, unlensed SMGs, and line emitters at similar redshifts. 
We also draw two types of relations for intrinsic CO line luminosity versus CO linewidth: one from \cite{Harris12} with $\mu=1$ and $\mu=21$, and the other from \cite{Bothwell13} with $\mu=1$ and $\mu=5.9$. 
The calculated magnification rates of J0107a from these relations, 21 and 5.9, respectively, differ by a factor of more than 3. 
{One reason for  this large difference is that both of these relations are empirical, and affected by the difference of the sample.
The other reason is the uncertainty in the calculation from \cite{Bothwell13}.
In this calculation, the parameter $C$, which parameterize the kinematics of the galaxy \citep{Bothwell13}, depends on the galaxy's mass distribution and velocity field, and it takes a wide range from $C\leq1$ to $C\geq5$ \citep{Erb06}. We here adopt $C=2.1$, as suggested in \cite{Bothwell13}, but it may include large uncertainty.}
In our case, the magnification rate from \cite{Harris12} is reasonable considering the consistency with the ratio of $(M_{\rm{gas}}+M_{\star})/M_{\rm{dyn}}$.

Such a high magnification must result in a highly perturbed morphology of the magnified image \citep[e.g.,][]{Tamura15}. In fact, we can find a hint of an elongated arc-like structure seen in the 0\farcs2 resolution 887 $\mu$m continuum image {(see Figure \ref{fig:J0107a} (a13))}, although the presence of a lens source is unclear because of contamination by the nearby bright source VV 114. 
{This is also true in the images of }\it{}Spitzer\rm{}{/IRAC and }\it{Chandra}\rm{}{/ACIS-I, which are shown in Figure \ref{fig:HST} (a and b). } 
From this point of view, the intrinsic radius of J0107a might be smaller than that we adopt here. 
In this case, $M_{\rm{dyn}}$ of J0107a would become even smaller than the result in Section \ref{subsec:dynamical mass}. Further, deeper, and higher-angular-resolution ALMA observations of this source will clarify the presence of a strong lens and the true size of J0107a.

\section{Conclusion} \label{sec:conclusion}

We conducted analysis of ALMA band 3, 4, 6, and 7 data of J0107a and J0107b, which are serendipitously discovered millimeter-wave line-emitting galaxies in the same field of the nearby galaxy VV114. 
In addition, we performed kinematic modeling of J0107a and investigated its physical properties. Our findings and conclusions are as follows:\\

1. We identify three emission lines, CO(4--3), CO(3--2), and [C\,{\sc i}](1--0), for each of them. In addition, we detect dust continuum emission of J0107a in bands 4, 6, and 7, and that of J0107b in band 6 { and 7}. \\

2. By fitting the Gaussian to CO spectra, we 
derive the redshifts of J0107a and J0107b as $z=2.4666\pm0.0002$ and $z=2.3100\pm0.0002$, respectively.\\

3. We obtain $M_{\rm{gas}}$ of our targets with the use of [C\,{\sc i}] line fluxes as $M_{\rm{gas}}(\mathrm{J0107a})=(11.2\pm3.1)\times10^{10}\ M_{\odot}$ and $M_{\rm{gas}}(\mathrm{J0107b})=(4.2\pm1.2)\times10^{10}\ M_{\odot}$, respectively. 
Moreover, {using $\alpha_{\rm{CO}}=0.8$ \citep{Downes98} and CO(1--0) luminosity which is converted from CO(4--3), we derive $M_{\rm{gas}}$ independently as $M_{\rm{gas}}(\mathrm{J0107a})=(13.4\pm1.7)\times10^{10}\ \ M_{\odot}$ and  $M_{\rm{gas}}(\mathrm{J0107b})=(5.0\pm0.7)\times10^{10}\ \ M_{\odot}$, respectively. 
These results are consistent with $M_{\rm{gas}}$ derived with [C\,{\sc i}] line fluxes within uncertainties.} \\

4. {We also calculate $M_{\mathrm{gas}}$ {of our targets with the dust continuum emission in 1.3 mm as $M_{\rm{gas}}(\mathrm{J0107a})=(3.2\pm1.6)\times10^{11}\ M_{\odot}$ and $M_{\rm{gas}}(\mathrm{J0107b})=(9.6\pm4.8)\times10^{10}\ M_{\odot}$,} respectively. These $M_{\rm{gas}}(\mathrm{dust})$ is consistent with $M_{\rm{gas}}$, which is derived with [C\,{\sc i}] line intensity, within uncertainties.} \\

5. {We make the rotating disk model of J0107a with GalPaK$^{\rm{3D}}$. This model reproduces not only the moment maps of J0107a, but also the position-velocity diagram of it well. This result suggests that J0107a is likely to have a rotating disk.} Using the results of this kinematic modeling, we obtain the dynamical mass of J0107a as $M_{\rm{dyn}}=(1.1\pm0.2)\times10^{10}\ M_{\odot}$. \\

6. We utilize the results of kinematic modeling to calculate the  ratio  of maximum rotation velocity $v_{\rm{max}}$ to velocity dispersion $\sigma_v$, namely $v_{\rm{max}}/\sigma_v$, and derive $v_{\rm{max}}/\sigma_v=1.3\pm0.3$.
The  $v_{\rm{max}}/\sigma_v$ of J0107a is quantitatively comparable to that of SMGs at similar redshifts.
\\

7. By comparing $M_{\rm{dyn}}$ and the sum of $M_{\rm{gas}}$ and $M_{\star}$, 
we propose that J0107a is magnified by the gravitational lens effect, and detected fluxes might be magnified by a factor of more than 10. 
This is consistent with the excess CO luminosity of J0107a compared to the expectation from the CO linewidth.

\begin{acknowledgments}
We thank the anonymous referee for a number of insightful comments and suggestions that greatly improved the quality of this paper.
We also thank Hideki Umehata and Ken Tadaki for helpful discussions on stellar properties of dust-obscured starburst galaxies. 
Data analysis was partly carried out on the common-use data analysis computer system at
the Astronomy Data Center (ADC) of the National Astronomical Observatory of Japan (NAOJ).
This paper makes use of the following ALMA data: ADS/JAO.ALMA\#2013.1.01057.S, \#2015.1.00973.S, \#2015.1.00902.S, and \#2013.1.00740.S. ALMA is a partnership of ESO (representing its member states), NSF (USA) and NINS (Japan), together with NRC (Canada), MOST and ASIAA (Taiwan), and KASI (Republic of Korea), in cooperation with the Republic of Chile.
The Joint ALMA Observatory is operated by ESO, AUI/NRAO and NAOJ. 
This work was supported by JSPS KAKENHI Grant Number JP17H06130 and by the NAOJ ALMA Scientific Research Grant Number 2017-06B. 
FE is supported by JSPS KAKENHI
Grant Number 17K14259.
DE acknowledges support from a Beatriz Galindo senior fellowship (BG20/00224) from the Ministry of Science and Innovation.
\end{acknowledgments}


\appendix
\section{cross-correlations between the GaLPak$^{\mathrm{3D}}$ results}

\begin{figure*}      
\gridline{\fig{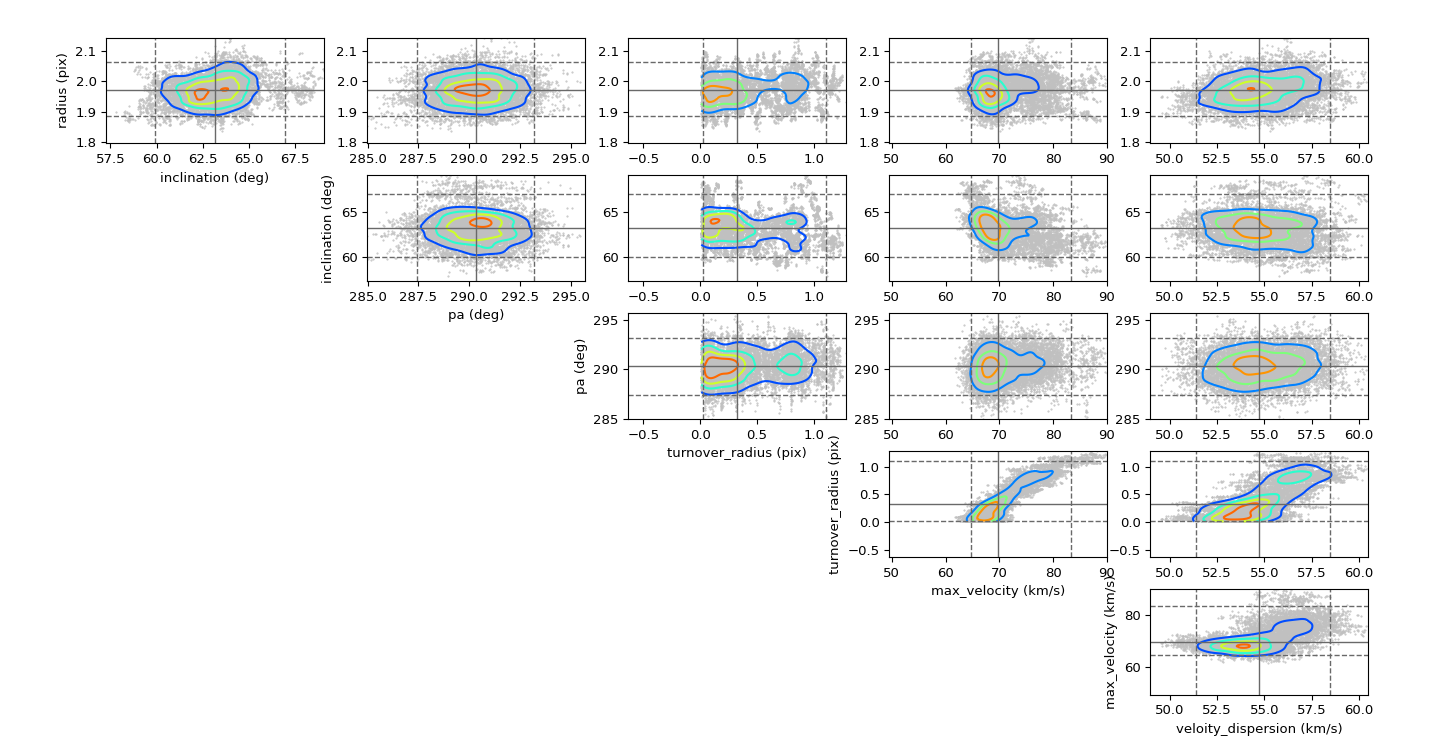}{1.0\textwidth}{}
          }
\caption{The cross-correlation plot of Markov chain in GalPaK$^{\rm{3D}}$ modeling with initial inclination angle of $i=60\tcdegree$. We use the last 10,000 iteration data of total 20,000 iterations to avoid using the data before MCMC burn-in. The gray dots indicate 
          individual iteration data and the density of these data is presented with contours. The solid and dashed lines indicate the average (50th percentile) and 2--$\sigma$ width (2.5th and 97.5th percentiles) of each parameter, respectively.}
\label{fig:correlation}
\end{figure*}

\bibliography{J0107}{}
\bibliographystyle{aasjournal}

\end{document}